\documentstyle[12pt]{article}

\def\sepand{\rule{14cm}{0pt}\and}
\newcommand{\beq}{\begin{equation}}
\newcommand{\eeq}{\end{equation}}
\newcommand{\bq}{\begin{quotation}}
\newcommand{\eq}{\end{quotation}}
\newcommand{\bc}{\begin{center}}
\newcommand{\ec}{\end{center}}
\newcommand{\BFACE}[1] {\mbox{\boldmath $#1$} }

\begin{document}

\title{Maupertuis principle, Wheeler's superspace and
an invariant criterion for local instability in general
relativity}

\vspace{0.6 cm}

\author{
{\sc Marek Biesiada} \\
{\sl Nicolaus Copernicus Astronomical Center, } \\
{\sl Bartycka 18, 00-716 Warsaw, Poland}  \\
\sepand
{\sc Svend E. Rugh} \\
{\sl The Niels Bohr Institute } \\
{\sl Blegdamsvej 17, 2100 K\o benhavn \O, Denmark}  \\
}

\date{}

\maketitle
\vfill
\begin{abstract}
\noindent
It is tempting to raise the issue of (metric) chaos in general
relativity since the Einstein equations are a set of highly
nonlinear equations which may exhibit dynamically very
complicated solutions for the space-time metric.
However, in general relativity it is
not easy to construct indicators of chaos
which are gauge-invariant.
Therefore it is reasonable to start by investigating
- at first - the possibility of a gauge-invariant
description of local instability.

In this paper we examine an approach which aims at describing the
dynamics in purely geometrical terms.
The dynamics is formulated as a geodesic flow
through the Maupertuis principle and
a criterion for local instability of the
trajectories may be set up in terms of
curvature invariants (e.g. the Ricci scalar) of the manifold on which
geodesic flow is generated. We discuss the relation of
such a criterion
for local instability (negativity of the Ricci scalar)
to a more standard criterion for local instability
and we emphasize that no inferences
can be made about global chaotic behavior from such local criteria.

We demonstrate that the Maupertuis principle implemented in the case
of the Hamiltonian (ADM) formulation of General Relativity is equivalent
to the construction of Wheeler's superspace on which the
development of three-metrics occurs along geodesics.
Obstructions for using
the superspace metric as a distance measure on the space of three-metrics
are pointed out. The discussion is illustrated in the case of
a particular toy-model of a gravitational collapse.

\end{abstract}
\vfill

\newpage

\section{Introduction}

Due to the strong nonlinearity of Einstein's equations one is tempted
to raise the issue of chaos in general relativistic context
especially in situations where this nonlinearity is particularly pronounced
- for example near a generic gravitational
collapse to a space-time singularity.
The property of chaos is usually understood in terms of (exponential)
divergence in {\em ``time''} of
{\em ``nearby orbits''} starting from slightly
modified initial conditions.
However one faces in the context of general relativity
some fundamental problems which make the
straightforward application of the machinery developed in the theory of
non-linear dynamical systems difficult.
At first, general relativity, i.e. Einstein's theory of gravitation,
is not even a dynamical theory in the usual sense.
It does not from the very beginning provide us
with a set of parameters (describing the gravitational degrees of freedom)
evolving in  ``time''.
Time looses here its absolute meaning as opposed to the classical dynamical
theories where the Newtonian time is taken for granted and upon which all our
experience in dealing with chaos is based. The division between space and time
in general relativity comes through foliating the space-time manifold $\cal M$
into spacelike hypersurfaces $\Sigma_t$. The metric $g_{\mu \nu}$ on $\cal M$
induces a metric $g_{ij}$ on $\Sigma_t$ (the first fundamental form of
$\Sigma_t$) and can be parametrized in the form
$$
g_{\mu \nu} = \left(
\begin{array}{cc}
N_i N^i - N^2 & N_j \\
N_i & g_{ij}
\end{array}
\right)
$$
where $N$ and $N_i$ are called lapse function and shift vector respectively.
Only after splitting the space-time into
space and time (the 3+1 ADM splitting) \cite{MTW,ADM}
we yield the possibility of
mapping the Einstein equations to a dynamical system
which resembles ordinary Hamiltonian
dynamical systems. In general this $3+1$ split is
quite an arbitrary procedure
and there is a priori  {\em no preferred time} coordinate.

The question we shall address in this paper is in particular the following:
\begin{center}
{\em What is meant by a ``nearby orbit'' in general relativity?}
\end{center}
Let us suppose that we have
performed a $3 + 1$ split of the spacetime metric into a three-metric
``${}^{(3)} g$'' evolving in a time coordinate ``$t$''. What is
then a
natural distance measure $  || {}^{(3)} g - {}^{(3)} g^{*} || $
between two three-metrics ${}^{(3)} g$ and ${}^{(3)} g^{*}$ at some given
moment $t$ of time?
In the case of non-relativistic Hamiltonian dynamical systems,
a standard indicator of ``chaos'' such like a principal Lyapunov
exponent (cf., e.g., discussion in \cite{EckmannRuelle,Lichtenberg,Benettin})
is calculated by using an {\em Euclidean} distance
measure which is naturally induced from the structure of the kinetic energy
term appearing in the non-relativistic Hamiltonian.
In general relativity it is, however, not obvious why we should use an
Euclidean distance measure
to assign a distance
between two space-time metrics.

In the class of spatially homogeneous spacetime metrics
\footnote{For a review of spatially homogeneous space-time metrics,
see e.g. \cite{Ryan,MacCallum}}
\begin{equation}  \label{spacetimehomogeneous}
ds^2 = - N^2 dt^2 + \gamma_{ij} (t) \BFACE{\omega}^i (x)
\BFACE{\omega}^{j} (x)
\end{equation}
calculations
of principal (maximal) Lyapunov exponents
have previously been reported
\cite{Burd,Hobill,Rugh1990a,RughJones}
for the dynamically interesting
case of the vacuum mixmaster gravitational collapse.
In fact, the mixmaster toy-model collapse
is a useful laboratory to test ideas about how to characterize
chaos in general relativity (see
also e.g. discussions in Rugh \cite{Rugh94}).
In the case of the mixmaster metric  $\BFACE{\omega}^i (x)$
in (\ref{spacetimehomogeneous}) denote the $SU(2)$ invariant one-forms
of the mixmaster space \cite{MTW}
and the three-metric is parametrized by three scale factors
$a = e^{\alpha}, \; b = e^{\beta}, \;c = e^{\gamma}$,
\begin{equation}
\gamma_{ij} (t) = diag(a^2 (t), b^2 (t), c^2 (t))
= diag(e^{2 \alpha} (t), e^{2 \beta} (t), e^{2 \gamma} (t)) \; .
\end{equation}
The Einstein equations for the mixmaster model are
given by three coupled non-linear second order
differential equations supplemented by a first
integral constraint (cf. e.g. Landau and Lifshitz \cite{LandauLifshitzII}).
It is tempting to treat the mixmaster collapse like
any other dynamical system (with few degrees of freedom),
integrate (numerically) the 6-dimensional state vector\footnote{Here
$\tau = \int dt/abc$ denotes the standard time variable in
Landau and Lifshitz \cite{LandauLifshitzII} and a
subscript $\tau$ means differentiation with respect to the time
variable $\tau$.}
$(\alpha, \beta, \gamma, \alpha_{\tau}, \beta_{\tau},
\gamma_{\tau}) (\tau) $
and extract ``Lyapunov exponents''\footnote{We note that
the spectrum of Lyapunov exponents is one of the most
well known (and often calculated) standard indicators of chaos in
a dynamical system.} in a standard way.
In all previous studies \cite{Burd,Hobill,Rugh1990a,RughJones}
Lyapunov exponents were
calculated by used an Euclidean distance measure
(between two ``nearby'' solutions) of the form\footnote{
Preferably, one measures the Euclidean distance constructed
from the entire state vector
$(\alpha, \beta, \gamma,
\alpha', \beta', \gamma')$
in phase space. What
is equivalent - Lyapunov exponents are
calculated from the
Jacobian matrix
corresponding to the state vector
$(\alpha, \beta, \gamma,
\alpha', \beta', \gamma')$
and integrating up along the orbit. Cf. e.g.
Hobill \cite{Hobill} or appendix A in S.E.Rugh
\cite{Rugh1990a}.}
\begin{equation} \label{EuclidianIntroduction}
|| \delta {}^{(3)} g_{ij} ||^2 =
(\delta \alpha)^2 + (\delta \beta)^2 + (\delta \gamma)^2
\end{equation}
Such choice of a distance measure is directly inspired
from the study of ordinary (non-relativistic) Hamiltonian
systems. In general relativity there is however a priori
no reason why one should use such a distance measure
to give out the distance between ${}^{(3)} g$ and the
``nearby'' three-metric $ {}^{(3)} g + \delta {}^{(3)} g$.
This distance measure has been put in artificially in hand and is
not supported by the structure of general relativity.
A distance measure which is induced by the
structure of the (ADM) Hamiltonian formulation of
general relativity is
\begin{equation} \label{LorentzianIntroduction}
|| \delta {}^{(3)} g_{ij} ||^2 = ds^2 = G_{AB} \delta g^{A} \delta g^B =
G_{ijkl} \delta g^{ij} \delta g^{kl}
\end{equation}
where $G_{AB} = G_{ijkl}$ is the so-called ``(mini)superspace'' metric.
We will here critically
examine the idea of using the ``superspace'' metric
$G_{AB}$ (or a conformally rescaled supermetric)
as a distance measure between two three-metrics.
In particular we discuss the idea of
extracting an invariant from it like the Ricci scalar in order
to give invariant (local) statements about the divergence of
``nearby'' orbits.
Thus, in this paper we
examine and extend in various ways
the discussion of an approach
\cite{BieSzy89} - \cite{Bie94}
which has proposed
a criterion for instability of the mixmaster
gravitational collapse. Szyd{\l}owski and {\L}apeta \cite{SzyLap},
Szyd{\l}owski and Biesiada \cite{BieSzy91}
proposed to look at the
Ricci scalar of the manifold on which the mixmaster dynamics generates
a geodesic flow
and investigate if one could extract
(coordinate invariant) information about instability properties and chaos
of the gravitational collapse in this way.

We show in sec.2.2 that for the case of non-relativistic Hamiltonian dynamics
the instability criterion of negativity of the Ricci scalar (of the
manifold on which the dynamics generates a geodesic flow)
is related to a more standard local instability criterion, which
examines local instability in terms of the eigenvalues
of the Jacobian of the Hamiltonian flow.
We will stress the
nontrivial interplay between local and global instability
criteria (since several authors draw false conclusions from local instability
criteria). In particular, we do not believe in expressions
directly relating the average negative curvature to  Lyapunov exponents
(see, e.g., Szyd{\l}owski and {\L}apeta \cite{SzyLap}
and Szyd{\l}owski and Biesiada \cite{BieSzy91}).

We note in sec.3 that application of the Maupertuis principle to
the Hamiltonian formulation of the mixmaster dynamics
- and the associated
induction of a natural distance measure on the three-metrics -
is exactly to implement the
dream by Wheeler, DeWitt and others
(described e.g. in \cite{Misner72} for the mixmaster metric)
to have a superspace metric
``$G_{ijkl}$'' (Wheeler's superspace) -
with respect to which the three-metrics move along geodesics -
and use this as a natural distance measure on the space of
mixmaster three-geometries.
In a sense such a superspace metric
would be the most natural to impose on the configuration space
of three-geometries (and appear
more natural than introducing a completely arbitrary Euclidean
metric on the space of three-geometries - as has been done
previously in calculating Lyapunov exponents for the
mixmaster collapse dynamics).

Thus, the Ricci scalar calculated in
Szyd{\l}owski and Biesiada \cite{BieSzy91}
from the Hamiltonian introduced by Bogoyavlenskii \cite{Bogoyavlenskii}
is {\em exactly} to calculate the Ricci scalar of
Wheeler's superspace metric
and the criterion $R < 0$ imply that we have
local instability at least in one direction when we look at the
geodesic deviation equation for two nearby mixmaster metrics
investigated with Wheeler's superspace metric as distance measure.

An obstacle to
this approach is the accompanying introduction
of a host of singularities of the superspace metric
which makes the dream of Wheeler
troublesome to achieve (see also discussion in C.W. Misner
\cite{Misner72}) - even in the
restricted class of mixmaster three-geometries. Invariants calculated from
the superspace metric, e.g. the Ricci scalar, inherit the singularities
of the Wheeler superspace metric. In particular, the Ricci scalar
calculated in Szyd{\l}owski and Biesiada \cite{BieSzy91}
has such singularities.
This obstacle has also most recently been emphasized
by Burd and Tavakol \cite{BurdTavakol}. However, as a {\em local}
instability criterion, $R < 0$ indicates - at
nonsingular points where it is defined
- the {\em local} exponential instability in some directions of
the configuration space.

As a major obstacle, we emphasize that the distance measure induced by
using the Wheeler superspace metric is not
positive definite.
The situation is somewhat similar to the indefiniteness
of the line element in special relativity (zero distances along
the light-cone).
But in the case of Wheeler's superspace metric in
general relativity there is nothing which prevents a trajectory
(a three-metric) from crossing the null-surface.
 Thus, one may contemplate situations where
a ``zero'' distance (between two three-metrics which need not be
identical) evolves into a finite (positive or negative)
distance. This corresponds formally to a
(local) ``Lyapunov exponent'' which is ``$\infty$''.

\vspace{1.0 cm}

The investigation is to be viewed in the light of the following question:
Can one assign an invariant
meaning to ``chaos'' in the general relativistic context?

There is a problem of
transferring standard indicators of chaos, e.g.
the spectrum of Lyapunov exponents, to the general relativistic context,
since they are highly gauge dependent objects. This fact was pointed
to and emphasized in Rugh \cite{Rugh1990a,Rugh1990b}  - and was also
discussed in Pullin \cite{Pullin}. (See also more recent
discussions in Rugh \cite{Rugh94}).

One should try to develop indicators of chaos
which capture chaotic properties of the gravitational field
(``metric chaos'') in a way which is invariant under spacetime
diffeomorphisms - or prove that this can not be done!
It may very well be that there is a ``no go'' theorem,
for the most ambitious task of constructing
a generalization of a Lyapunov exponent
(extracted from the continuous evolution equations) which is
meaningful and invariant under the
full class of spacetime diffeomorphisms
(H.B. Nielsen and S.E. Rugh).
One should therefore, at first, try to modify the intentions and
construct instability measures which are invariant under a
smaller class of diffeomorphisms, e.g. coordinate transformations
on the spacelike hypersurfaces $\Sigma_t$ which do not involve
transformations of the time coordinate.

``Per time'' indicators - like a Lyapunov exponent -
have only a little chance of remaining {\em invariant} under
diffeomorphisms. Even if we consider the restricted class of
Lorentz transformations, i.e. space-time transformations within
the special theory of relativity, it appears, that the
time evolution of a dynamical system is slowed down
if we observe it
from a frame of reference ${\cal S}'$,
which moves (is boosted) with some velocity $v$ relative
to the frame of reference ${\cal S}$ in which the dynamical
system is at rest. Hence, the Lorentz $\gamma$-factor will
inevitably appear
in the Lyapunov exponent (seen from the moving frame of
reference ${\cal S}'$).
One notes, however, that we are able to discriminate qualitatively,
in any frame of reference, whether or not the
dynamical system is chaotic: The requirement that a
Lyapunov exponent is greater than zero
will be valid in any frame of reference if it is
valid in one. That is, ``$\lambda > 0$'' is a statement which is
{\em invariant} under Lorentz-transformations - although the numerical
value of the Lyapunov exponent $\lambda$ is not.
In general relativity, however, ``$\lambda > 0$'' is not
an invariant statement under diffeomorphisms more general
than Lorentz-transformations \cite{Rugh1990a,RughJones,Rugh1990b,Pullin}.

A route of progress may lie in pointing to indicators of purely
geometrical nature (``fractal dimensions'', etc).
We shall here explore into a picture which is purely geometrical
and stems from an old construction known since times of Maupertuis and Jacobi
--- namely that in which Hamiltonian dynamical systems are mapped into geodesic
flows.
In this approach the entire complicated dynamics comes out
from purely geometrical
properties of a single ``object'' --- the manifold on which
the geodesic flow is generated.
One may consequently dream that also such basic
properties like integrability or chaos may be somehow captured in purely
geometrical (and hence timeless and invariant) terms.
For example one may hope to describe local instability of trajectories in terms
of curvature invariants (e.g. the Ricci scalar) since it is curvature
that determines the behavior of close geodesics (through the geodesic deviation
equation). We shall however also point out several obstructions one
has to face when implementing such a line of thinking
to general relativistic problems.

\section{The dynamics as geodesic motion by application of Maupertuis
principle}

\setcounter{equation}{0}

It is indeed a remarkable fact that a large class of
Hamiltonian
systems can be presented in such a way that Hamilton's equations
acquire formal resemblance to
geodesic equations in Riemannian geometry.

The phase space trajectories
$\BFACE{x}(t) = (\BFACE{p}(t),\BFACE{q}(t))$ with respect to time
parameter $t$ and corresponding to the
Hamiltonian
\begin{equation}  \label{OriginalHamiltonian}
H = H(\BFACE{p},\BFACE{q}) =\sum_{i,j} \frac{1}{2} a^{ij} p_i p_j +
V(\BFACE{q})
\end{equation}
may be mapped into a geodesic motion on
the configuration space manifold.
It is accomplished in a way which
displays a complete analogy with the mapping of the
evolution of three-metrics into that of geodesic motion in
Wheeler's superspace (see next section).

The equations of
motion for the $\BFACE{q}$ coordinates may be written as
\begin{equation} \label{Newton2law}
\ddot q^j
+ \tilde \Gamma^j_{\,ks}\,\dot q^s\,\dot q^k = - a^{ji}
 \frac{\partial V(\BFACE{q})}{\partial q^i}
\end{equation}
where $\tilde \Gamma^j_{\,ks}$ is the Christoffel symbol calculated with
respect to $a_{ij}$ metric.
Due to the force term this is, obviously, not a geodesic equation.
It is simply the Newton's second law restated.
The  momentum variables are
just linear combinations of
velocities $p_i = a_{ij} \dot q^j$.
Transformation to a geodesic motion (i.e. free motion in a curved space)
is accomplished in two steps:
$(1)$ conformal transformation of the metric $a_{ij}$, and
$(2)$ change of the time parameter along the orbit. In fact only the
first step is crucial the second one merely introduces the affine parameter
along the geodesics.
More explicitly we equip the configuration space
with the (super)metric
\begin{equation}  \label{SupermetricMaupertius}
g_{ij} = 2 (E - V(\BFACE{q})) a_{ij}
\end{equation}
(note that $a_{ij}$ is read off from the kinetic energy term in the
Hamiltonian and (in general) is allowed to vary as a function of
the configuration space variable, $a_{ij} = a_{ij}(\BFACE{q}))$
). \\
With respect to the metric (\ref{SupermetricMaupertius})
and the time parameter $t$ it is not easy to see
that the orbits are geodesics since there is a
term appearing on the right hand side
of the equation,
\begin{equation} \label{notgeodesicMaupertius}
\displaystyle{d^2\over dt^2}\,q^i  +
\Gamma^i_{\,jk}\,\displaystyle{d\over dt}
q^j\,\displaystyle{d\over dt}q^k = - \frac{1}{E - V(\BFACE{q})}
\frac{d}{dt}q^i \frac{\partial}{\partial q^k} V(\BFACE{q})\frac{d}{dt} q^k
\end{equation}
where $\Gamma^i_{\,jk}$ now denote the  Christoffel symbols
associated with the (super)metric (\ref{SupermetricMaupertius}).
However, if we (re)parametrize the orbit $q^i = q^i(s)$
in terms of the parameter $s$ defined via
\begin{equation}  \label{parameterMaupertius}
s = \int 2\;(E-V)\;d \; t
\end{equation}
the orbits will become affinely parametrized geodesics\footnote{Let us
recall from differential
geometry that the most general form
of the geodesic equation (when the geodesic is parametrized by $v$ parameter)
reads $q^{''i} + \Gamma^i_{jk} q^{'j} q^{'k} = h(v) q^{'i}$ (prime denotes
differentiation with respect to $v$) where $h(v)$ is an arbitrary function of
$v$. Then by solving the differential equation $\frac{d^2v}{ds^2} +
h(v){\frac{dv}{ds}}^2 = 0 $ one can construct a new parameter
--- the so called affine parameter $s$ such that
$\ddot q^i + \Gamma^i_{jk} \dot q^j \dot q^k = 0$ (overdot here denotes
differentiation with respect to $s$). It is evident that this new
parameter is defined up to linear transformations $s \to A\,s + B$
($A,B $ are constants) what justifies its name.}, i.e. the
configuration space variables $\BFACE{q} = (q^i)$ satisfy the
well known geodesic equation
\begin{equation} \label{geodesicMaupertius}
\displaystyle{d^2\over ds^2}\,q^i + \Gamma^i_{\,jk}\,\displaystyle{d\over ds}
q^j\,\displaystyle{d\over ds}q^k = 0
\end{equation}
with no force term on the right hand side.

There is no ``magic'' in this, it is merely a
rewriting of the Hamiltonian equations.
The information about the original force acting on the particle
(as described by the potential $V(\BFACE{q})$ in the Hamiltonian
(\ref{OriginalHamiltonian}))
has been encoded entirely  in the definition of the (super)metric
(\ref{SupermetricMaupertius})
and the definition of the new parameter $s$ in
(\ref{parameterMaupertius}) parametrizing the orbit.
\footnote{Note that the above picture comes quite naturally from the
Maupertuis-Jacobi least action principle
$$\delta S = \delta \;\int_{q'}^{q''} \sqrt{E-V(\BFACE{q})}
\sqrt{a_{ij}\;dq^idq^j} = 0 $$
and its formal resemblance to the variational formulation of geodesics in
Riemannian geometry as curves extremalizing the distance.
We have sketched the step-by-step derivation of the geodesic
motion in order to see the step-by-step analogy with the derivation of
three-metrics as geodesics in Wheeler's superspace in the next section.}
The entire procedure
involves the following principal steps:

\begin{quotation}
\noindent
{\bf 1.} At first a metric $a_{ij}$ is read off from structure of kinetic
term in the Hamiltonian.  \\
{\bf 2.} We observe that the trajectory $\left\{ \BFACE{q} (t)
\right\} $ is not
a geodesic with respect to this metric. \\
{\bf 3.} Transformation of the non-geodesic motion
to affinely parametrized geodesic motion is obtained via two steps: \\
{\bf (1)} Conformal transformation of the metric $a_{ij} \rightarrow
2(E-V(\BFACE{q}))a_{ij} $  \\
{\bf (2)} Rescaling of the time parameter
$t \rightarrow s = \int 2 (E - V(\BFACE{q})) dt $
\end{quotation}

It is, however, not often that this description is used
in the discussion of Hamiltonian dynamical systems,
for example in the discussion of
instability properties and
chaotic behavior of such systems.
We contemplate several reasons why this  approach is
so rarely used:

\begin{quotation}
\noindent
{\bf 1.} The metric $g_{ij}$ in
(\ref{SupermetricMaupertius}) and the
parameter $s$
in (\ref{parameterMaupertius}) are singular in ``turning'' points
where
the kinetic energy term $\sum \frac{1}{2} a^{ij}\,p_ip_i$ is zero and we thus
have
$$ E - V(\BFACE{q}) = 0 \; . $$
Therefore we have to consider the motion
away from the ``turning'' points.
This may be a more severe restriction
for systems with
few degrees of freedom, e.g. the harmonic oscillator with a single degree
of freedom, than for systems with many
degrees of freedom, e.g. for  N self-gravitating bodies
in a gravitational field, to
which the virial theorem applies, and thus
$\sum \frac{1}{2} a^{ij}p_ip_j = 0$ is rather unlikely.

\noindent
{\bf 2.} The procedure of mapping to geodesic motion (via the Maupertuis
principle)
is applicable to {\em any} time independent Hamiltonian system
of the form (\ref{OriginalHamiltonian}) where the potential $V(\BFACE{q})$ is
not allowed
to depend on momenta $\BFACE{p}$. I.e. the application of Maupertuis
principle is not straightforward if we for example
have a mechanical system coupled to an electromagnetic field with the
Hamiltonian
$$ H = \frac{1}{2m} (\BFACE{p}-\frac{e}{c}A(\BFACE{q}))^2 + V(\BFACE{q}).$$\\
The same objection holds for mechanical problems in noninertial frames, for
example the problem of motion of a star in the field of a rotating galaxy
\cite{Geneva}.
\end{quotation}

In spite of these principal limitations of the Maupertuis approach, it has
been explored in various contexts.
In his seminal work \cite{Krylov} on foundations of statistical mechanics
Krylov has suggested that viewing the dynamics of
an N-body system as a geodesic
flow on the appropriate manifold may provide a universal (applicable to a very
large class of interaction potentials) tool for discussing relaxation
processes. Inspired by works of Hedlund and Hopf \cite{HedlundHopf} Krylov
conjectured that the
{\em average} separation of trajectories evolves according to
the sign of the Ricci
scalar $R$ on the configuration space accessible for the system.
An implementation of this viewpoint have also been
attempted in the context of general relativity
by Szyd{\l}owski, Biesiada et al.
\cite{BieSzy89,SzyLap,BieSzy91}.
As we shall demonstrate, Wheeler's original
dream to construct a superspace
in which the three metrics move {\em along geodesics}
is {\em exactly} to
implement the old idea of Maupertuis principle in general relativistic
context. We shall however first - in the next two subsections -
comment on the status of the $R<0$ criterion and its relation
to more standard criteria for local instability.

\subsection{The Ricci scalar {$R$} for the ``super'' metric
and  {$R < 0$} as a (strong) local instability criterion.}

Contemplating the local instability properties,
i.e. how nearby orbits behave, it is
natural to consider the geodesic deviation equation
which describes the behavior of nearby geodesics (\ref{geodesicMaupertius}).

This can be derived in the usual manner by subtracting the
equations for the geodesics $q^i(s)$ and $q^i(s) + \xi^i (s)$ respectively
or simply by disturbing the fiducial trajectory $(p_i(t),q^i(t))$,
\begin{eqnarray}
\tilde{p}_i(t) & = & p_i(t) + \eta_i(t), \nonumber \\
\tilde{q}^i(t) &= & q^i(t) + \xi^i(t)
\end{eqnarray}
and substituting this directly into Hamilton's equations.
In this way we also
arrive momentarily, though tediously,
at the geodesic deviation equation for the configuration space
variable $\BFACE{\xi}$ (the separation vector),
\begin{equation} \label{geodesicdeviationMaupertius}
\frac{D^2\,\xi^i}{D\,s^2}=-R^i_{jkl}\,u^j \xi^k u^l
\end{equation}
Here $\BFACE{u}= D\BFACE{q}/ Ds $ is the tangent vector  to
the geodesic, $\BFACE{\xi}$ is the
separation vector orthogonal to $\BFACE{u}$.
Note, that the covariant
derivative $D / Ds $ and Christoffel symbols are calculated
with respect to the (super)metric (\ref{SupermetricMaupertius}). This
(super)metric induces the natural distance measure
$$ || \BFACE{\xi} ||^2 = g_{ij} \xi^i \xi^j =
2(E - V(\BFACE{q})) a_{ij} \xi^i \xi^j $$
on the configuration space.
 From the geodesic deviation equation (\ref{geodesicdeviationMaupertius}) one
arrives at the inequality (see e.g. \cite{Bie94})
\begin{equation} \label{inequalityK}
\frac{d^2}{ds^2} ||\BFACE{\xi}||^2 \geq - 2  K_{u;\xi}
||\BFACE{\xi}||^2
\end{equation}
where the $K_{u;\xi}$
denotes the sectional curvature in the two-direction $\BFACE{u} \wedge
\BFACE{\xi}$ . \\
Noting that the Ricci scalar may be expressed as a sum of the principal
sectional curvatures,\footnote{The principal sectional curvatures
are defined, as is well known, as the extrema of
the sectional curvature $K_{u;\xi}$ as function of the wedge product
$\BFACE{u} \wedge \BFACE{\xi}$.}
\begin{equation}
R = Tr \; \BFACE{K} =
\sum\;
\left(
\begin{array}{c}
Principal \;Sectional \\
\; \;  Curvatures
\end{array}
\right)
\end{equation}
the inequality (\ref{inequalityK}) especially implies that if
\begin{equation}
R < 0
\end{equation}
then a principal direction
$\xi$ exists along which we have local exponential instability.
\footnote{Note that if  there is one direction $\BFACE{\xi}$
(for a given
$\BFACE{u}$)
in which there is exponential instability then there is also
a continuum of nearby exponentially unstable trajectories.
Due to smoothness of
$K_{u;\xi}$
as a function of two-direction the
sectional curvature for nearby directions $\xi$ is also negative.}

By first calculating the Christoffel symbols $\Gamma_{ij}^{k} $
from the metric (\ref{SupermetricMaupertius}) and subsequently substituting
this into the standard formula for the Ricci scalar $R $ one obtains
the following general expression,
\begin{equation} \label{RMaupertius}
R (Maupertuis)  =  \frac{(n-1)}{2(E-V)^3} \sum_{i,j = 1}^{n} \left\{
(E-V) \frac{\partial^2 V}{\partial q^i \partial q^j} a^{ij}
- \frac{(n-6)}{4}
\frac{\partial V}{\partial q^i} \frac{\partial V}{\partial q^j}
a^{ij} \right\}
\end{equation}
corresponding to a Hamiltonian system with
$n$ degrees of freedom.

N.S. Krylov \cite{Krylov} was - as far as we know - the first
to emphasize the use of
$R < 0$ as an instability criterion
in the context of an $N$ body system (a gas) interacting via Van der
Waals forces, with the ultimate hope to understand
the relaxation processes in a gas.
In his toy-model study he found that indeed $R < 0$ in
the accessible domain of the configuration space and concluded that this
provides an explanation for the relaxation as a consequence of dynamical
mixing.\footnote{The investigations by Krylov
was strongly influenced by seminal results of
Hedlund and Hopf \cite{HedlundHopf} where the geodesic motion on the
Lobachevsky space was proved to be mixing. Whereas the approach by
Krylov is intuitively appealing it is however not allowed to
transfer the results obtained for the Lobachevsky space to the case where
sectional curvatures are negative only in average and not at
every point and in every two-direction $\BFACE{u} \wedge \BFACE{\xi}$.
Moreover, Krylov disregarded the problem of compactness of the
configuration space manifold which is important for making
inferences about mixing.}
Another investigation in the same spirit, have been performed by Gurzadyan
and Savvidy \cite{GurzadyanSavvidy} for the case of
collisionless stars interacting via gravitational forces, also
with the purpose of understanding the relaxation
properties of such a stellar system.
\footnote{Physical
motivation of this study comes from the observation that
stellar systems (globular clusters and galaxies) are apparently in
a well relaxed state that is reflected in regularity of their shapes,
velocity dispersions, surface luminocities etc. On the other hand
the most obvious relaxation process -- namely via binary encounters,
provides a relaxation time that is greater than the Hubble time
\cite{Chandrasekhar2}.
As a possible way out of this puzzle Lynden-Bell \cite{Lynden-Bell}
has proposed the so called violent relaxation mechanism.
Gurzadyan and Savvidy attempted at looking at the relaxation process
in stellar systems from Krylov's viewpoint. The  idea was that
if they could show that a  self-gravitating N-body system was a K-system
it would indicate that the system approached the equilibrium exponentially
fast. }

The system of collisionless stars
which was investigated by V.Gurzadyan and G.K. Savvidy \cite{GurzadyanSavvidy}
and Kandrup
\cite{Kandrup} had $R < 0$
whenever the number of stars in the system exceeded two but
{\em not} on a domain which was {\em compact}.
Kandrup \cite{Kandrup}, however,
argues that the
self-gravitating system is ``effectively bound'' i.e.
the phase-space is
compact at least for sufficiently small times. If one waits long enough
the effects of ``evaporation'' of stars from the system come into play,
so the phase-space is actually infinite.
\footnote{Had the manifold been compact it would - of course -
still not be possible to deduce that the system is a K-system.
The K-system property holds for geodesic flows on compact manifolds
of negative curvature, i.e. manifolds whose sectional curvature
$K_{u; \xi}$ is negative at {\em every} point and in
{\em every} two-direction $\BFACE{u} \wedge \BFACE{\xi}$.
On the other
hand it is a general property of Hamiltonian systems that sufficiently close to
the physical boundary of motion there exist both positive and negative
sectional curvatures \cite{vanVelsen}. This means, in particular,
that mapping a Hamiltonian flow to a geodesic flow via Maupertuis principle
can never be used to show that the system is a K-system.
We note, however, that to be a K-system is a
very strong, highly non-generic property
to satisfy.
If a system is ``less'' than a K-system, it may
- of course - nevertheless be ``highly'' chaotic (and unstable almost
all over in the phase space). }
Let us emphasize that the negativity of the Ricci scalar
provides only a sufficient condition for {\em local} instability
and it is thus not possible
to make any rigorous statements about chaos from local instability
properties
alone as it will be discussed in more detail in the next subsection.

\subsection{Remarks on the instability criterion $R < 0$ and its
relation to  more ``standard'' instability criteria.}

According to Szebehely \cite{Szebehely} there are more than fifty
distinct criteria of instability!
Many of these are related to each other
(though not all connections are worked out in a transparent
way). Some of the criteria are even in conflict with each other.

Let us emphasize a relation between two important local criteria, the
criterion of having a negative Ricci scalar $R = R(Maupertuis) < 0$
(corresponding to the formulation via the Maupertuis
principle) and the more usually applied criteria connected
to the eigenvalues of the Jacobian of a Hamiltonian flow.

$R < 0$  is a {\em sufficient} criterion for {\em local} instability ---
but it is not a necessary criterion
and very often it fails to say anything of interest about
our dynamical system. For instance, in
the case of a simple class of Hamiltonian flows
\begin{equation}  \label{xyHarmonic}
H = E = \frac{1}{2} (p_x^2 + p_y^2) +
\frac{1}{2} (x^2 + y^2) + \frac{1}{2} x^2 y^2
\end{equation}
it fails to mirror the transition
from almost integrable to almost completely chaotic (almost K-system)
as the energy $E$
increases.
A calculation of the Ricci scalar, according to formula
(\ref{RMaupertius}) or (\ref{RMaupertiustwo})
shows that it is always positive irrespective of the
Energy E (since $\bigtriangleup V(\BFACE{q}) = 2 + x^2 + y^2 > 0$).
Despite this, the model is
almost a K-system in the limit for large energies (i.e. when the nonlinear
term $\frac{1}{2} x^2 y^2$ becomes very large), see discussions
in P. Dahlqvist and G. Russberg
\cite{DahlqvistRussberg} and references therein.

A standard criterion \footnote{In an attempt to
predict the onset of dynamical chaos
(a key ingredient of which is the exponential instability of adjacent
trajectories)
Toda \cite{Toda} and Duff and Brumer \cite{Brumer} proposed
the criterion based on evaluation of the Gaussian curvature of the
potential $V(\BFACE{q})$.}
for local instability
is to look at the eigenvalues of
the Jacobian of the Hamiltonian flow.
(However, wrong conclusions as regards global chaos are often drawn from it.)
The line of thinking starts at disturbing the fiducial trajectory
$(p_i(t),q^i(t))$:
\begin{eqnarray} \label{fiducial}
\hat p_i(t) & = & p_i(t) + \eta_i(t), \nonumber \\
\hat q^i(t) &= & q^i(t) + \xi^i(t)
\end{eqnarray}
and subsequently investigating the evolution of the disturbance vectors
$\xi$ and $\eta$. The stability of
motion is determined
by integrating along the trajectory (see later) the
time dependent
matrix (the Jacobian matrix of the Hamiltonian flow):
\begin{equation}
\BFACE J =
\left( \matrix{\BFACE 0 & \BFACE 1\cr
- Hess(V({\BFACE q}(t))) & \BFACE 0 \cr }\right)
= \left( \matrix{\BFACE 0 & \BFACE 1\cr
-(\frac{\partial^2 V({\BFACE q}) }{\partial q_i \partial q_j})_{ij} (t)
& \BFACE 0 \cr }\right)
\end{equation}
where the Hessian of the potential $V(\BFACE q)$ is evaluated along the
reference trajectory making the $\BFACE J$-matrix time dependent.
The inconvenience of the time dependence
can be overcome by replacing the
time-dependent phase-point $\BFACE{q}(t)$ (moving
along the trajectory) by a fixed phase-space coordinate
$\BFACE{q}$ resulting in
\begin{eqnarray} \label{Variation}
\dot \xi^i & = &  \eta^i = a^{ij} \eta_j\\
\dot \eta_i & = &  - \sum_{ij} \frac{\partial^2 V(\BFACE{q})}{\partial q^i
\partial q^j} \xi^j \nonumber
\end{eqnarray}

This step is sometimes claimed ``mathematically dubious'' (cf. also, e.g.,
M.Tabor \cite{Tabor}).
Anyway, the $\BFACE{J}$-matrix is now time-independent and
the stability of the autonomous system (\ref{Variation})
is determined from the eigenvalue problem
$$ \det \left[ \BFACE{J} - \lambda \BFACE{1} \right] = 0$$

If at least one eigenvalue has a positive real part we have exponential
growth (locally) of the disturbance vector in one direction of the phase
space.
For example, in the case of a 2-dimensional Hamiltonian we have the
two pairs of roots
$$
\lambda_{\pm} = \pm  \frac{1}{\sqrt{2}} \left\{
- (\frac{\partial^2 V}{\partial^2 q^1} + \frac{\partial^2 V}{\partial^2 q^2})
\pm \sqrt{
(\frac{\partial^2 V}{\partial^2 q^1} + \frac{\partial^2 V}{\partial^2
q^2})^2 - 4
(\frac{\partial^2 V}{\partial^2 q^1} \frac{\partial^2 V}{\partial^2 q^2}
- (\frac{\partial^2 V}{\partial q^1 \partial q^2})^2)
} \right\}^{1/2}
$$
In particular, if
$ det(\frac{\partial^2 V}{\partial q^i \partial q^j}) =
\frac{\partial^2 V}{\partial^2 q^1} \frac{\partial^2 V}{\partial^2 q^2}
- (\frac{\partial^2 V}{\partial q^1 \partial q^2})^2 < 0$ (i.e. negative
Gaussian curvature of the potential $V$)
a pair of roots becomes
real and we have local instability.

The $R < 0$ criterion implies the J-instability above\footnote{This
is not surprising from simple considerations: If we have
instability along some
direction in the configuration space
(as captured by the $R <0$ criterion)
then we also have instability at least
in one direction in phase space. }
(whereas the converse is not true). In
this sense the $R < 0$  is a
{\em stronger instability criterion} than the J-instability.
For example, for the two dimensional Hamiltonian system (with
$a_{ij} = \delta_{ij}$)
the Ricci scalar (\ref{RMaupertius}) reads:
\begin{eqnarray}  \label{RMaupertiustwo}
R (Maupertuis) & = &
- \frac{1}{2 (E - V(\BFACE{q}) )}  \bigtriangleup
\ln (E - V(\BFACE{q}) )  \; \nonumber \\
& = & \;  \frac{1}{2 (E - V(\BFACE{q}) )^2 }    \;
( \bigtriangleup V(\BFACE{q}) \; + \;
\frac{ (\bigtriangledown V(\BFACE{q}) )^2 }{E - V(\BFACE{q}) }  \; )
\end{eqnarray}
where $\bigtriangleup = \sum_{i} \partial^2 / \partial (q^i)^2$
is the Laplacian (if $a_{ij}=\delta_{ij}$).\\
Thus from $R<0$ follows that $\bigtriangleup V < 0$ which in turn
implies the
existence of eigenvalues $\lambda$ with a positive real part
irrespective of the sign of the Gaussian curvature of the potential.
On the other hand the J-instability criterion
(sometimes called the Toda-Brumer criterion)
may capture the change of local instability properties of
the Hamiltonian flow even in cases when
$\bigtriangleup V > 0$ and hence $R>0$ .
For example, the instability criterion $R < 0$ fails to capture the
transition from KAM-integrability to chaos (as the energy $E$
increases) in the class of
Hamiltonian flows (\ref{xyHarmonic}), whereas the study of the Jacobian
of the flow (the Toda-Brumer-Duff criterion) gives a threshold
value $E = E_c$ (onset of ``local instability'') for the energy.
Thus, a small calculation will verify that for energies
$$ E > E_c = 3/2 $$
there are eigenvalues
of the Jacobian $\BFACE{J} (x) $
with a positive real part.
Let us emphasize, what is well known,
that it is not possible to make conclusions about chaos
from this local criterion.
Global instability (chaos) is defined by looking at the spectrum of
Lyapunov exponents:
Consider the set of first order differential equations,
$ \dot{\vec{\BFACE{x}}} = \vec{\BFACE{f}}(\vec{\BFACE{x}}), \;
\vec{\BFACE{x}} = (\BFACE{p}, \BFACE{q}) \in \BFACE{R}^{n} $
(the Hamiltonian flow).
The Jacobian $\BFACE{J}$ of the flow
$\vec{\BFACE{f}}$ is the $n \times n$ matrix
$\BFACE{J} = (\partial \vec{\BFACE{f}} / \partial \vec{\BFACE{x}})$.
In order to define the Lyapunov spectrum of characteristic exponents
connected to {\em a given} trajectory $\vec{\BFACE{x}}(t)$
with initial conditions $\vec{\BFACE{x}}(0)$
one first define the stability matrix,
\begin{equation} \label{stabilitymatrix}
\BFACE{M} = M_{ij}(\vec{\BFACE{x}}(0),t) \equiv
\frac{\partial x_i(t)}{\partial x_j (0)}
\end{equation}
which satisfies the differential equation
\begin{equation}
\dot{\BFACE{M}}_{ij} =
\BFACE{J}(\vec{\BFACE{x}}(t))_{ik} \BFACE{M}_{kj}
\; ,\; \; \BFACE{M}(0)=\BFACE{1}
\end{equation}
The solution to this differential equation is called the exponential
of the Jacobian under the flow, symbolically denoted by
\begin{equation}  \label{Jexponentiated}
\BFACE{M}(t) =
\hat{T} \exp (\int_{0}^{t} \BFACE{J}(\vec{\BFACE{x}}(t)) dt )
\BFACE{M}(0)
\end{equation}
where $\hat{T}$ is the time ordering operator (the $\BFACE{J}$'s do not
commute, being $n \times n$ matrices).\footnote{
$\hat{T} \exp (\int_{0}^{t} \BFACE{J}(\vec{\BFACE{x}}(t)) dt )$
may be calculated as
$\hat{T} \lim_{\Delta t \rightarrow 0}
\prod (\BFACE{1} + \BFACE{J}(\vec{\BFACE{x}}(t) \Delta t )
\BFACE{M}(0)  $
the product being taken along the integrated trajectory
$\left\{ \vec{\BFACE{x}}(t) \right\} $ in time
steps $\Delta t \rightarrow 0$.}
In order to determine whether a trajectory is {\em stable} (regular)
or not, one investigates {\em the growth} of
the stability matrix
$\BFACE{M} = \BFACE{M}(\vec{\BFACE{x}}(0), t )
= \BFACE{M}(t) $ with time.
More explicitly, consider
$
\Lambda (t) = ( \BFACE{M}^{ \dagger} (t) \BFACE{M} (t) )^{1/2}
$
(cf., e.g., Eckmann and Ruelle, \cite{EckmannRuelle} p. 630).
Let the $n$ eigenvalues of the $n \times n$ matrix $\Lambda (t)$
(which are real and positive) be denoted
$ d_1 (t) \geq d_2(t) \geq ... \geq d_n (t)$.
The {\em Lyapunov functions} are defined as
$
\lambda_j(t) = t^{-1} \log d_j(t)
$
and the associated {\em spectrum of
Lyapunov exponents} \footnote{Theoretical criteria for
{\em convergence} of this limit,
which holds under rather general assumptions of the
flow $\vec{\BFACE{f}}$,  are discussed in e.g.
Eckmann and Ruelle \cite{EckmannRuelle}.
The {\em maximal}
characteristic exponent $\lambda_{max}$
is the easiest to find. But this is
also sufficient to detect the exponential divergence of
nearby orbits, called ``sensitive dependence on
initial conditions''.
Per definition, a dynamical system displays a stochastic chaotic behavior
in a region of the phase space {\em if} its maximal Lyapunov characteristic
exponent is positive for trajectories in this region, cf.
Benettin et al. \cite{Benettin}. },
\begin{equation}
\lambda_j = \lim_{t \rightarrow \infty} \lambda_j (t) =
\lim_{t \rightarrow \infty} \left\{ \frac{1}{t} \log d_j (t) \right\} \; \; .
\end{equation}
In the very special case where the Jacobian is constant in time,
$\BFACE{J} (\vec{\BFACE{x}}(t)) = \BFACE{J}_0 $
we get $\BFACE{M} (t) = \exp (\BFACE{J}_0 \cdot t)$.
If $\lambda$ is an eigenvalue of $\BFACE{J}_0$ we have
that $e^{\lambda t}$ is an
eigenvalue of $\BFACE{M} (t)$. It follows,
that $\lambda$ is one of the Lyapunov exponents and local
J-instability is equivalent with global instability (Lyapunov
exponents).

In general, however, no implication exists between local and global
instabilities. This is illustrated by a simple model example.
Consider, e.g., a two-dimensional map
$$ x \rightarrow (\BFACE{A} \BFACE{B} \BFACE{A} \BFACE{B}
\BFACE{A} \BFACE{B} \BFACE{A} \BFACE{B} ...) x \; \; , \; \;
x \in \BFACE{R}^2 $$
where a $2 \times 2$ matrix $\BFACE{A}$ is multiplied on the vector
every second time, and the $2 \times 2$ matrix
$\BFACE{B}$ every second time.

If we choose
$\BFACE{A} = \left( \matrix{1 & \lambda \cr
0 & 1 \cr }\right) $ and
$\BFACE{B} = \left( \matrix{1 & 0 \cr
\lambda & 1 \cr }\right) $, $\lambda \in \BFACE{R}$,
both are locally stable, since the eigenvalues are equal to 1.
$\BFACE{AB}$ has eigenvalues $1 \pm \lambda$ and globally the
system is unstable. On the other hand if we choose
$\BFACE{A} = \left( \matrix{\Lambda & 0 \cr
0 & \lambda \cr }\right) $ and
$\BFACE{B} = \BFACE{A}^{-1} = \left( \matrix{1/\Lambda & 0 \cr
0 & 1/\lambda \cr }\right) $, where $\Lambda > 1$ and $0 < \lambda < 1$,
both $\BFACE{A}$ and $\BFACE{B}$ gives local instability while globally the
system is stable ($\BFACE{AB} = \BFACE{1}$).

The same non-trivial relationship between local
and global instability criteria holds for continuous Hamiltonian
flows:
Orbits $\gamma$ may build up a ``global instability'' (as captured
by the exponentiated Jacobian
along the orbit, cf. formula (\ref{Jexponentiated}))
despite they are stable {\em everywhere} in the phase space
with respect to the local instability criterion.
For example, if we consider the simple
Hamiltonian toy-model (\ref{xyHarmonic}),
for energies $E < E_c = 3/2$ there are still
many orbits which are globally unstable, despite
the fact that there are no positive, real eigenvalues of the Jacobian
in any accessible phase space point for the trajectory.

Moreover, we may have (globally)
stable orbits (Dahlqvist and Russberg \cite{DahlqvistRussberg})
which are everywhere locally unstable.
(Consider the Hamiltonian (\ref{xyHarmonic})
in the limit $E \rightarrow \infty$.
The Jacobian has real, positive eigenvalues in any point of
the accessible phase space. Nevertheless,
there is an orbit of period 11 which is
globally stable, cf. Dahlqvist and Russberg \cite{DahlqvistRussberg}).

Thus there is a non trivial interplay between the local instability (as
mirrored in the eigenvalues of the stability matrix
(\ref{stabilitymatrix})) and the global,
exponentiated Jacobian (\ref{Jexponentiated}),
yielding the ``Lyapunov exponents'' along the orbit.
These remarks are well known. We repeat
them here to comment on what one may
conclude (or not conclude)
from local criteria of instability like
$$ R(Maupertuis) < 0. $$
There has been some confusion concerning this point in the past.

Especially, we can not from $R < 0$ (local instability)
deduce that a Lyapunov exponent is positive (global instability).

As already noticed in the introduction there were attempts
\cite{SzyLap,BieSzy91,SzSzczBie} to
relate the $R<0$ criterion with Lyapunov exponents.
In particular it was argued
that the quantity
\begin{equation} \label{RLyap}
\lambda = lim_{\tau \rightarrow \infty}
\sqrt{\frac{- R}{n(n-1)}}
\frac{1}{\tau}
\int_0^{\tau} 2(E - V(\BFACE{q}(t))) d t
\end{equation}
played a role of the
principal Lyapunov exponent. The Ricci scalar $R$ is meant here as an
average measure of local divergence of trajectories.
The motivation of such
claims (see \cite{SzyLap,BieSzy91,SzSzczBie}) had its roots in
the following facts: If we take the geodesic deviation equation
(\ref{geodesicdeviationMaupertius}) and rewrite it as a
gradient system (see appendix 1 in Arnold \cite{Arnold}), i.e.
$$
\frac{D^2 \BFACE{\xi}}{Ds^2} = - {\BFACE{\nabla}}_{\xi}[V_u(\xi)] =
-{\BFACE{\nabla}}_{\xi}(\frac{1}{2} R_{ijkl} \xi^i u^j \xi^k u^l) =
- {\BFACE{\nabla}}_{\xi} \; [\displaystyle \frac{1}{2}
K_{u;\xi} g(\xi,\xi)] $$
where the ``potential'' $V_u(\xi)$ has been  expressed by sectional curvature:
$V_u(\xi) =
\displaystyle \frac{1}{2} K_{u;\xi} g(\xi,\xi)$.
Then by averaging the geodesic deviation equation over the orientation of the
geodesic (i.e. at a given point we take all possible geodesics passing
through this point, average over the bivector $\BFACE{u}
\wedge \BFACE{\xi} $ and denote this average by
$ < \; > $) one arrives at the
formula \cite{SzSzczBie}:
$$
\frac{D^2 <\xi^i>}{Ds^2} = - \frac{R}{n(n-1)} <\xi^i>,
$$
from which one may argue for
the quantity (\ref{RLyap}) bearing in mind that the
$s$-parameter is related to the $t$-time by (\ref{parameterMaupertius}).
The reasoning is flawed, however, in several aspects.
One cannot replace the integration of a full Jacobian matrix,
cf. equation (\ref{Jexponentiated}),
with the integration of a Ricci-scalar which is just some average
quantity  (trace over the sectional curvatures).
The negativity of the Ricci scalar, $R<0$,
which is an initial assumption for (\ref{RLyap}) to apply is only a sufficient
criterion of local instability.
Hence in the case
of the Hamiltonian (\ref{xyHarmonic}), for example, we have $R > 0$
whereas the Lyapunov exponent $\lambda > 0$ for most trajectories.
Then apart from
the fact that local instability cannot be simply translated into global chaos
also the meaning of ``averaged deviation equation'' is questionable.
Even if the Ricci
scalar $R$ is traced (numerically) along trajectories one may give examples of
systems with positive Kolmogorov entropy
(understood, here, as the sum over positive Lyapunov exponents)
for
which the quantity (\ref{RMaupertius}) is either positive or negative
(see e.g. discussion in \cite{Geneva}).
The above critique apply to some earlier papers
\cite{SzyLap,BieSzy91,SzSzczBie} and also to Burd and Tavakol who adopted
the expression (\ref{RLyap}) in \cite{BurdTavakol}.

We would also like to question the validity of the
formula (38) in Gurzadyan and Savvidy
\cite{GurzadyanSavvidy} in which
the Ricci scalar is related to
a relaxation time of the dynamical system
consisting of collisionless stars interacting via gravitational
forces.

In order not to close this section in too pessimistic mood let us stress that
the Maupertuis-Jacobi procedure of reducing the dynamics
to geodesic flows is a
powerful tool for geometrizing the dynamical systems\footnote{
The geometrical setting arising from the Maupertuis-Jacobi principle
makes this picture suited for investigating integrability of Einstein's
equations as a consequence of existence of Killing vectors and tensors in the
minisuperspace \cite{Uggla}.
Also, the fact that the Maupertuis principle involves
variations at a fixed energy makes this formulation useful for the
microcanonical ensemble and proves useful in deriving thermodynamical
properties of gravitating systems as advocated by Brown and York
\cite{BrownYork}.
These possible applications of the Maupertuis principle are
beyond the scope of the present paper and
the interested reader is referred to a
review paper \cite{Geneva} and the references therein.}
(provided one is bearing
in mind its limitations, see also \cite{Geneva}).

A serious problem in general relativity
is to define a metric space in which our spacetime metric
(which is itself a metric space) is a phase space point and
where the divergence of trajectories makes sense.
The next section shall reveal the intimate connection between such metric
structure called the superspace and the Maupertuis-Jacobi principle implemented
to the General Relativity in its Hamiltonian (ADM) formulation.

\section{Wheeler's superspace: Maupertuis principle
implemented in the context of general relativity.}

\setcounter{equation}{0}

An original dream by
Wheeler, DeWitt and others is to consider the dynamics of the three
metrics ${}^{(3)} g$ as geodesics on some manifold
called superspace\footnote{Abstractly the
``superspace'' is meant to be
the space of all three-metrics
modulo diffeomorphisms on the three-space, i.e. the quotient space
${\cal S}({\cal M}) = Riem({\cal M})/Diff({\cal M})$.
}
equipped with the metric tensor
$ G_{ijkl}$  which is named the
supermetric. This supermetric $G_{ijkl}$ induces a norm
on the space of three metrics ${}^{(3)} g $,
\begin{equation}
|| \delta g_{ab} ||^2 = \int d^3 x \sqrt{g} \;
G^{ijkl} \delta g_{ij} \delta g_{kl}
\end{equation}
and one may measure the distance between two three-metrics
${}^{(3)} g$ and ${}^{(3)} \tilde{g}$ with respect to
the $G_{ijkl}$ tensor.  (In practice it may be difficult to find
such global distances. Like in the case of finding distances between
Copenhagen and Warszawa, say, one has to minimize over all
paths connecting the given two points).

What can one choose for the distance measure $G_{ijkl}$ ?
At first there are many possible  choices of metrics on the
space of three metrics. For example (cf. DeWitt \cite{DeWitt})
one could choose a metric like
\begin{equation}
G^{ijkl} = \frac{1}{2} (g^{ik}g^{jl} + g^{il}g^{jk})
\; \delta (x, x')
\end{equation}
which has the property that the distance between two three-metrics
is zero if and only if the three-metrics are identical.
However, another distance measure (supermetric) is suggested
from the ADM Hamiltonian formulation of general relativity.
It is read off from the kinetic term in the Hamiltonian constraint,
\begin{equation}
\pi_i^k \pi_k^i - \frac{1}{2} (\pi_k^k)^2 - g {}^{(3)} R = 0
\end{equation}
(cf. e.g. Misner \cite{Misner72} and references therein) associated with the
constrained Hamiltonian formalism  which may be set up
for the Einstein equations (see e.g. discussion and references in Teitelboim
\cite{Teitelboim}).
More explicitly,
$$ \pi_k^i \pi_i^k - \frac{1}{2} ( \pi_k^k)^2 =
\frac{1}{2}(g^{ik} g^{jl} \pi_{ij} \pi_{kl} + g^{il}g^{jk} \pi_{ij} \pi_{kl}
- g^{ij} g^{kl} \pi_{ij} \pi_{kl}) = G^{ijkl} \pi_{ij} \pi_{kl} $$
and we thus arrive at
\begin{equation}
G^{ijkl} = \frac{1}{2}(g^{ik}g^{jl} + g^{il}g^{jk} - g^{ij}g^{kl}).
\end{equation}

Note that the supermetric, in this notation
differs by a conformal factor ($\sqrt{g}$)
from DeWitt's expression  \cite{DeWitt}.
This is allowed since the ADM action is invariant with respect to
conformal transformations.

Let us now see how we may arrive at the geodesic equation for
the three-metric ${}^{(3)} g$. It turns out that it can be done in a way which
is completely analogous, step-by-step,
to the previous section dealing with the similar question for an
ordinary (nonrelativistic)
Hamiltonian system $ H = \frac{1}{2} a_{ij} p^i p^j + V(q)$.

The procedure  involves the following principal steps:
\begin{quotation}
\noindent
{\bf 1.} The metric $G_{AB}$ is read off from the ADM Hamiltonian. \\
{\bf 2.} Observe that we do not have geodesic equation for
$\left\{ {}^3 g \right\}$,
with respect to metric $G_{AB}.$ \\
{\bf 3.} Transformation to (affinely parametrized)
geodesic motion of $\left\{ {}^3 g \right\}$ is obtained in two steps: \\
{\bf (1)} Conformal transformation of the metric to $\tilde{G}_{AB} =
{\cal R} G_{AB}.$  \\
{\bf (2)} Rescaling of the $\lambda$ parameter, $\lambda \rightarrow
\tilde{\lambda}$.
\end{quotation}

{\bf 1.} As we have already noted, from the structure of the Hamiltonian

\begin{equation}
H =\frac{1}{2} G_{(ij)(kl)}\pi^{ij} \pi^{kl} - \frac{1}{2} g {}^{(3)}R
=\frac{1}{2} G_{AB}\pi^{A} \pi^{B} - \frac{1}{2} g \; {}^{(3)}R = 0
\end{equation}
we read off the first candidate for a metric on the configuration space
(the space of three-metrics) \footnote{The
covariant form of the supermetric $G_{ijkl}$ actually differs from its
contravariant counterpart.
This is the consequence of a natural duality of these two tensors:
$$G^{ijmn}G_{mnkl} = \frac{1}{2} (\delta^i_k \delta^j_l + \delta^i_l \delta^j_k
)$$}
\begin{equation}
G_{AB} \equiv
G_{(ij)(kl)} = \frac{1}{2}(g_{ik}g_{jl} + g_{il}g_{jk}
- 2 g_{ij}g_{kl}).
\end{equation}
This step corresponds to reading off the ``metric'' $a^{ij}$  from
the Hamiltonian
$$H =\frac{1}{2} a^{ij}p_i p_j + V(\BFACE{q}) = E $$
in the previous section. \\

{\bf 2.} We observe that we do not have a geodesic equation for the
trajectory ${}^{(3)}g(\lambda)$ \footnote{By $\lambda$ we denote the parameter
which parametrizes the evolution of three-metrics --- the so-called supertime.
For more detailed discussion see C.W.Misner \cite{Misner72}}.
There is a ``force term'' on the right hand side:
\begin{equation} \label{parasupergeodesic}
\frac{d^2 g^A}{d \lambda^2} + \Gamma^{A}_{BC} \frac{dg^B}{d \lambda}
\frac{dg^C}{d \lambda} =
\frac{1}{2} G^{AB}  \frac{\partial {\cal R}}{\partial g^B} =
\frac{1}{2} G^{AB}  \frac{\partial (g {}^{(3)} R) }{\partial g^B}
\end{equation}
where
\begin{equation}
\Gamma^A_{BC} = \frac{1}{2} G^{AD}(\frac{\partial G_{BD}}{\partial g^C}
+ \frac{\partial G_{DC}}{\partial g^B}
- \frac{\partial G_{BC}}{\partial g^D} )
\end{equation}
Equation (\ref{parasupergeodesic})
may be one-to-one translated to the similar expression
(\ref{Newton2law}) from
sec.2,
where we also have a force term,
$- a^{ji} \partial V(\BFACE{q})/ \partial q^i$, in the form of  the metric
(read off from the kinetic term) multiplied by
the gradient of the potential. The
role of the potential in this general relativistic context is
played by the quantity
\begin{equation} \label{potentialgenrel}
V = - \frac{1}{2}g \; {}^{(3)}R \equiv - \frac{1}{2} {\cal R}   \; \; .
\end{equation}
In analogy with section 2 (Maupertuis principle) we want
to make the three-metrics move
along geodesics !    \\

\noindent

{\bf 3.} Transformation of the non-geodesic motion (above) of
${}^{(3)}g$ to geodesic motion is obtained in two steps:  \\

\noindent
(1) First we make a conformal transformation of the metric
\begin{equation} \label{conformalMisner}
\tilde{G}_{AB} = {\cal{R}} G_{AB} \; \; , \; \;
\tilde{G}^{AB} = \frac{1}{\cal{R}} G^{AB}
\end{equation}
This conformal transformation of the metric (\ref{conformalMisner})
is analogous to equation (\ref{SupermetricMaupertius}) in
section 2, i.e. the conformally rescaled metric,
$$ g_{ij} = 2(E - V(q)) a_{ij}  $$
since
\begin{equation}
\tilde{G}_{AB} = {\cal R} G_{AB} = (-2 V) G_{AB} =
2 (E - V) G_{AB}
\end{equation}
(note that the energy $E$ is zero in the general relativistic context
by virtue of the Hamiltonian constraint). \\
Analogously to sec. 2 conformal rescaling of the metric alone is not
sufficient to map the evolution of three-metrics into {\em affinely
parametrized} geodesics.\\

\noindent
(2) However, by rescaling of the $\lambda$ parameter,
\begin{equation}
d \tilde{\lambda} = 2(E - V) d \lambda =
- 2 V d \lambda = {\cal R} d \lambda =
g {}^{(3)} R \; d \lambda
\end{equation}
we obtain now the important result (cf. also e.g. Misner \cite{Misner72})
that with respect to this new parameter $\tilde{\lambda}$
and the conformally rescaled
metric $\tilde{G}_{AB}$ the
three-metric ${}^{(3)} g$ is now an affinely parametrized geodesic (in this
``Superspace'')
\begin{equation}  \label{WheelerSupergeodesic}
\frac{d^2 g^A}{d \tilde{\lambda}^{2}} +
\tilde{\Gamma}^{A}_{BC} \frac{dg^B}{d \tilde{\lambda}}
\frac{dg^C}{d \tilde{\lambda}} = 0.
\end{equation}

Like in the previous section
there is no ``magic'' involved (we have ``magic without magic'').
Information about the potential
${}^{(3)}R$
has simply been encoded completely in the mathematical definition
of the Superspace-metric $\tilde{G}$
with respect to which the evolution of ${}^{(3)}g$ then becomes
geodesic motion (affinely parametrized if one uses the rescaled parameter
$\tilde{\lambda}$).

Note, that instead of Misner's analogy with a free particle
in special relativity, with Hamiltonian $H = \frac{1}{2}(\eta^{\mu \nu}
p_{\mu} p_{\nu} + m^2)$ (cf. Misner \cite{Misner72}, p. 451) we have rather
emphasized here the complete one-to-one correspondence between construction of
the Superspace and dynamics of a non-relativistic
particle with the Hamiltonian $H = \frac{1}{2} a^{ij} p_i p_j +
V(\BFACE{q})$ reduced to geodesic flow by virtue of the Maupertuis principle.

Whereas we would not be surprised if the analogy above is
to some extent well known for those who understand the construction of
Wheeler's superspace, we have not seen the analogy stated so clearly
and it puts the previous investigations
\cite{BieSzy89,SzyLap,BieSzy91,SzSzczBie,Bie94}
along this route in a somewhat new perspective.
Namely they can be viewed as exploring the
geometric structure of the superspace aimed at investigating
local instabilities in the evolution of three-geometries.
Dreaming of this {\em purely geometrical} picture of dynamics one has
nevertheless to face several problems:

{\bf 1.} There may exist (and in fact
it is not of rare occurrence) points where the
potential term ${\cal R}$ is zero. At such points the Supermetric
(\ref{conformalMisner}) as well as the new parameter $\tilde{\lambda} = \int
{\cal R} d\lambda $
are not well defined and present at first an obstruction to this
beautiful idea of mapping the evolution of three-metrics into geodesic motion
w.r.t. this Superspace. Of course as already mentioned in sec.2 it is
merely an artifact which stems from the conformal transformation
(\ref{conformalMisner}) and nothing singular happens at those points to the
dynamics. This fact has been well known already to Misner
\cite{Misner72}. It has also
recently been emphasized by Burd and Tavakol \cite{BurdTavakol}.

{\bf 2.}
If we are to uphold the dream in the strongest version of having the
three-metrics as geodesics in superspace it should be the
conformally rescaled ``super'' metric
$\tilde{G}_{AB}$ which should be called the ``super metric''
(since that
is the one which generates geodesic motion of the ${}^3 g$)
and not the $G_{AB} = G_{ijkl} $ metric, which is more frequently
used as a ``superspace metric''.
This imply, however, that the
(mini)superspace metric
would differ, for example, from one minisuperspace model
of some Bianchi type to a minisuperspace model of another
Bianchi type - since the conformal rescaling  depends on the
specific space-time metric considered.

{\bf 3.}
The Supermetric  $\tilde{G}_{AB}$  induces a natural distance
measure on the three-geometries in the relevant superpace.
However, it should be noted, what is very well known,
that $\tilde{G}_{AB}$
represents a pseudo-Riemannian geometry (with signature (-- +++++)),
therefore we may have distances $ds^2$ which are
space-like, null or time-like.
(This indefiniteness of the distance measure is particular to
the application of Maupertuis principle in general relativity and
the natural distance measure induced by this procedure. Application of
Maupertuis principle to non-relativistic Hamiltonian systems give
rise to positive definite distance measures).
Therefore in general relativity
we may have the unfamiliar situation that two three-metrics
which are at zero distance with respect to the Wheeler's superspace
metric $\tilde{G}_{AB}$ may evolve into a finite (positive or negative)
distance. This corresponds ``formally'' to a
(local) ``Lyapunov exponent'' which is ``$\infty$''.

 \vspace{1.0 cm}

In analogy to the non-relativistic case but bearing in mind all
the limitations we may contemplate
the behavior of close trajectories (mapped into geodesics) starting
from the geodesic deviation equation:
$$ \frac{D^2 \xi^A}{D \tilde{\lambda}^2} =
- R^A_{BCD} \frac{d g^B}{d \tilde{\lambda}} \xi^C \frac{d g^D}
{d \tilde{\lambda}} $$
where $\xi= \delta {}^{(3)} g$ is the deviation vector between three-metrics,
and the Riemann Christoffel tensor $R^A_{BCD}$ is
calculated with respect to the supermetric $G_{AB}$.

The idea is to use
$$ R = - \frac{(n-1)}{{\cal R}^3} \sum_{A,B}
\left\{ {\cal R} \frac{\partial^2 {\cal R}}{\partial g^A \partial g^B} G^{AB}
+  \frac{(n-6)}{4} \frac{\partial {\cal R}}{\partial g^A}
\frac{\partial {\cal R}}{\partial g^B} G^{AB} \right\} < 0 $$
as a geometrical (and hence
coordinate invariant) local instability criterion for the three-metrics
evolving according to the Einstein's equations.
 If $R < 0$ then there exist at least one direction of perturbing the
three-metric $g^A \rightarrow g^A  + \delta g^A$ so we have a local
exponential amplification of the metric perturbation in that direction -
measured with respect to the Wheeler's superspace metric as a distance
measure on the three-metrics.

In the next section we shall apply this in a concrete example of the
orbits of the mixmaster minisuperspace gravitational collapse.

\begin{table}
\begin{tabular}{|c|c|}
\hline
\multicolumn{2}{||c||}{{\em Supermetrics (Maupertuis principle)
and local instability criteria}} \\
\multicolumn{2}{||c||}{{\em in non-rel. systems and in
general relativity }} \\
\hline \hline
\begin{tabular}{c}
  \\
  A Hamiltonian for a  \\
non-relativistic \\
mechanical system \\
  \\
\end{tabular} &
\begin{tabular}{c}
  \\
ADM-Hamiltonian \\
in General Relativity \\
(say, the mixmaster collapse) \\
  \\
\end{tabular}
\\ \hline \hline
\begin{tabular}{c}
{\bf 1.} Read off the first candidate \\
of a supermetric ``$a_{ij}$'' from the structure  \\
of the kinetic term in the Hamiltonian  \\
$ H = \sum \frac{1}{2} a^{ij}p_i p_j + V(q) $ \\
\end{tabular}            &
\begin{tabular}{c}
{\bf 1.} Read off the first candidate \\
of a supermetric ``$G_{AB}$'' from the structure  \\
of the kinetic term in the ADM-Hamiltonian  \\
$ H = \sum \frac{1}{2} G_{AB} \pi^A \pi^B - \frac{1}{2} g \;{}^{(3)} R = 0 $ \\
\end{tabular}
\\ \hline
\begin{tabular}{c}
{\bf 2.} We observe that the trajectory $\left\{ q(t) \right\} $ \\
do not obey a geodesic equation of motion \\
w.r.t. metric $a^{ij}$: There is a force term \\
$ - a^{ij} \partial V(\BFACE{q}) / \partial q^i $    \\
on the right hand side of (\ref{Newton2law}). \\
\end{tabular}   &
\begin{tabular}{c}
{\bf 2.} We observe that the trajectory $\left\{ g^A \right\} $ \\
do not obey a geodesic equation of motion \\
w.r.t. metric $G^{AB}$: There is a force term \\
$ G^{AB} (\partial (\frac{1}{2} g {}^{(3)} R) / \partial g^A )$    \\
on the right hand side of (\ref{parasupergeodesic}). \\
\end{tabular}
\\ \hline
\begin{tabular}{c}
{\bf 3.} Transformation to geodesic motion  \\
is obtained in two steps: \\
{\bf (1)} Conformal transformation of the metric \\
$ a_{ij} \rightarrow 2(E - V(q)) a_{ij} $    \\
{\bf (2)} Rescaling of the time parameter \\
$ dt \rightarrow ds = 2 (E-V(q)) dt $  \\
and we obtain the equation (\ref{geodesicMaupertius}). \\
\end{tabular}   &
\begin{tabular}{c}
{\bf 3.} Transformation to geodesic motion  \\
is obtained in two steps: \\
{\bf (1)} Conformal transformation of the metric \\
$ \tilde{G}_{AB} = 2(E - V) G_{AB} = (g\;{}^{(3)}R) G_{AB}$    \\
{\bf (2)} Rescaling of the $\lambda$ parameter \\
$ d \tilde{\lambda} = 2 (E-V) d \lambda = (g\;{}^{(3)}R) d\lambda $ \\
and we obtain the equation (\ref{WheelerSupergeodesic}). \\
\end{tabular}
\\ \hline \hline
\begin{tabular}{c}
The Ricci scalar may be calculated \\
from the resulting ``supermetric'' \\
and the criterion  \\
$ R < 0 $ \\
is used as a local instability criterion \\
(N-body systems etc., Krylov etc.) \\
\end{tabular}  &
\begin{tabular}{c}
The Ricci scalar may be calculated \\
from the resulting ``supermetric'' \\
and the criterion  \\
$ R < 0 $ \\
has been  attempted as a \\
local instability criterion \\
in the context of general relativity \\
(Szyd{\l}owski, Biesiada etc.) \\
\end{tabular}
\\ \hline \hline
\end{tabular}
\caption[XYZ]{{\small  summarizing the analogy between the Maupertuis
principle for non-relativistic Hamiltonian systems and Wheeler's
superspace (Maupertuis principle) for general relativity. }
}
\end{table}

\newpage
\subsection{Application to the
mixmaster gravitational collapse}

For the mixmaster homogeneous toy-model collapse with the three-metric,
$$ {}^{(3)} g_{IX} = \gamma_{ij}(t)
\BFACE{\omega}^i (x) \BFACE{\omega}^j (x) $$
where
$\gamma_{ij}(t) = diag(a^2(t), b^2(t), c^2(t))$
and $a,b,c$ are the scale factors of the metric, we have in the
ADM-variables,
\begin{equation} \label{ADMvariables}
\left(
\begin{array}{c}
\Omega \\ \beta_+ \\ \beta_-
\end{array}
\right) = \left(
\begin{array}{ccc}
- \frac{1}{3} & - \frac{1}{3} & - \frac{1}{3} \\
\frac{1}{6} & \frac{1}{6} & -\frac{1}{3} \\
\frac{\sqrt{3}}{6} & - \frac{\sqrt{3}}{6} & 0
\end{array}
\right)
\left(
\begin{array}{c}
\ln{a} \\ \ln{b} \\ \ln{c}
\end{array}
\right)
\end{equation}
the following form of the Hamiltonian
 \begin{equation}
 {\cal H} = \frac{1}{2}(G^{AB}p_A p_B - {\cal R}) =
\frac{1}{2} (-p_{\Omega}^2 + p_{+}^2 + p_{-}^2 +
e^{-4 \Omega}(V(\beta_{+}, \beta_{-}) - 1))
\end{equation}
where the mixmaster three-curvature potential reads
\begin{eqnarray}
{\cal R} & = &
- e^{-4 \Omega}(V(\beta_+, \beta_-) - 1)  \\
& = & - e^{-4 \Omega} \bigl[
\frac{2}{3} e^{4 \beta_+} (cosh(4 \sqrt{3} \beta_-) -1) + \frac{1}{3}
e^{- 8 \beta_+} - \frac{4}{3} e^{-2 \beta_+} cosh(2 \sqrt{3} \beta_-)
\bigr] \nonumber
\end{eqnarray}

The (mini)superspace variables are thus $g^A = (g^{\Omega}, g^+, g^-)
= (\Omega, \beta_+, \beta_-)$ ($A,B=1,2,3$) and the momenta are
$p_A = G_{AB} dg^B/d \lambda$ giving $p_{\Omega} = - d\Omega/d \lambda \;,
p_{\beta_+} = d \beta_{+}/d \lambda \;, \; p_{\beta_-} =
d \beta_-/d \lambda $.
We immediately read off the first candidate for a ``super'' metric
from the kinetic term in the Hamiltonian
$$G_{AB} = G^{AB} = diag(-1,+1,+1).$$
The Christoffel symbol corresponding to this flat ``supermetric'' vanishes,
$\Gamma^A_{BC} = 0$, and the Hamiltonian equations translate into
$$ \frac{d^2 g^A}{d \lambda^2} = \frac{1}{2} G^{AB}
(\frac{\partial {\cal R}}{\partial g^B}) \; \; \; . $$
I.e. it is a kind of Newton'second law with non-vanishing
force term appearing on the right hand side,
$ d^2 \Omega/d \lambda^2 = - \frac{1}{2} \partial {\cal R}/\partial \Omega$,
$ d^2 \beta_+ /d \lambda^2 = \frac{1}{2} \partial {\cal R}/\partial \beta_+$,
$ d^2 \beta_-/d \lambda^2 =  \frac{1}{2} \partial {\cal R}/\partial \beta_-$.

However, the configuration space variables
$g^A = (\Omega, \beta_+, \beta_-)$ may be mapped to a
(affinely parametrized)
geodesic flow on
the configuration space manifold equipped with the (mini)superspace metric
\begin{equation}
\tilde{G}_{AB} = {\cal R} G_{AB} = - e^{4 \Omega} (V(\beta_+, \beta_-) -
1) diag(-1,1,1)
\end{equation}
with the parametrization
\begin{equation}
d \tilde{\lambda} = {\cal R} d \lambda =
- e^{4 \Omega}(V(\beta_+, \beta_-) - 1) d\lambda
\end{equation}
along the trajectory of the mixmaster collapse. Thus,
we have
\begin{equation}
\frac{d^2 g_{IX}^A}{d \tilde{\lambda}^{2}} +
\tilde{\Gamma}^{A}_{BC} \frac{dg_{IX}^B}{d \tilde{\lambda}}
\frac{dg_{IX}^C}{d \tilde{\lambda}} = 0
\end{equation}
This procedure introduces a host of singularities at points where
${\cal R}=0$ i.e. where $V=1$ (see also an interesting discussion in Misner
\cite{Misner72} pp. 453-454).

The fact, recently emphasized also by
Burd and Tavakol \cite{BurdTavakol}
that the singularities introduced by the supermetric
$G_{ijkl}$ prevents us from mapping the entire mixmaster collapse orbit
into one single, unbroken and simple geodesic orbit - is of course an
important obstacle to this approach and
was, in fact, previously emphasized by one of us (S.E.R.). \\
Despite these troubles, let us suppose that we
nevertheless carry out these
transformations at points in the mixmaster configuration space
$g^A = (\Omega, \beta_+, \beta_-)$ where the
superspace metric $\tilde{G}_{AB} = \tilde{G}_{ijkl}$ do
{\em not} introduce
artificial singularities, i.e. at those
points where {$V(\beta_+, \beta_-) \neq 1$}.
With this supermetric $\tilde{G}_{ijkl}$ one may thus construct
non-trivial intervals in which
the mixmaster metric is mapped into geodesics
(in Wheeler's (mini)superspace).
We shall start with some remarks on
the properties of the distance measure
between two nearby mixmaster metrics ${}^{(3)} g_{IX}$ and
$ {}^{(3)} g_{IX} + \delta {}^{(3)} g_{IX}$.
Let us first note, that the distance between $g^A = (\Omega, \beta_+, \beta_-)$
and  $g^A + \delta g^A =
(\Omega + \delta \Omega, \beta_+ + \delta \beta_+ ,
\beta_- + \delta \beta_-)$ is of Lorentzian signature.
This indefiniteness is a property of
the supermetric, both in its original form
\begin{equation}
|| \delta g^A ||^2  = ds^2 = G_{AB} \delta g^A \delta g^B
= -d \Omega^2 + d \beta_+^2 +
d \beta_-^2
\end{equation}
and in the conformally rescaled form
\begin{equation}
|| \delta \tilde{g}^A ||^2 =
d\tilde{s}^2 = \tilde{G}_{AB} \delta g^A \delta g^B =
- e^{-4 \Omega}(V(\beta_+, \beta_-) - 1) ( -d \Omega^2 + d \beta_+^2 +
d \beta_-^2).
\end{equation}
Thus the distance can take
positive, zero or negative values and one has the unfamiliar
situation (relative to positive definite Euclidean
distance measures implemented in the context
of non-relativistic Hamiltonian dynamical systems) - as
previously stressed - that two
configuration space points (mixmaster three-
metrics) which is not identical
may have zero distance with respect to these naturally induced
distance measures.

These distance measures have on the other hand the good property
of being invariant under canonical coordinate transformations.
Thus, if we have a change of coordinates (configuration space variables),
$g^A \rightarrow g^{*A}$ the distance measure
$ds^2 = G_{AB} \delta g^A \delta g^B $ is invariant
$$ || \delta g^A ||^2 = ds^2 = G_{AB} \delta g^{A} \delta g^{B} =
G_{AB}^* \delta g^{*A} \delta g^{*B} = || \delta g^{*A} ||^2 $$
under such transformations, since $G_{AB}$ transforms properly as a
tensor,
\begin{equation} \label{transformation}
G^{*}_{AB}
= \frac{\partial g^C}{\partial g^{*}_A}
\frac{\partial g^D}{\partial g^{*}_B} G_{CD}.
\end{equation}

One can easily see this by recalling that the coordinate transformation
$g^A \rightarrow g^{*A}$ induces the
canonical transformation of momenta
$$ p_A \to p_A^{*} = \frac{\partial g^B}{\partial g^{*A}}\;p_B \; .$$
Therefore the Hamiltonian $\cal H$ reads
$${\cal H} = \frac{1}{2}(G^{AB}p_A p_B - {\cal R}) =
\frac{1}{2}(G^{AB}\frac{\partial g^{*C}}{\partial g^A}
\frac{\partial g^{*D}}{\partial g^B}p^{*}_C p^{*}_D - {\cal R}) =
\frac{1}{2}(G^{*AB}p^{*}_A p^{*}_B - {\cal R}) $$
thus justifying claims of the formula (\ref{transformation}).   \\

A quantity which is invariant under a large class of coordinate
reparametrizations  (canonical transformations in the
Hamiltonian formulation)
was considered in works by Biesiada,
{\L}apeta, Szcz{\c e}sny and Szyd{\l}owski
\cite{SzyLap,BieSzy91,SzSzczBie}
where the Ricci scalar of the manifold on
which the mixmaster model acts as a geodesic flow has been extracted.
As we have seen here this is precisely to extract the Ricci scalar for the
Wheeler's superspace metric $\tilde{G}_{AB}$
(in the conformally rescaled version where it
describes the three-metrics
as geodesics) and
negative values $R < 0$ are naturally interpreted
as a local measure of
exponential instability via the geodesic deviation equation
with respect to the
conformally rescaled distance measure - of Lorentzian signature - on the
space of mixmaster three-metrics.

Despite the problems with unwanted
singularities induced by this approach (as noted by e.g. Misner
\cite{Misner72}) we believe that insufficient attention
has been paid towards the
use of distance measures naturally induced by the
structure of general relativity, e.g. the
Wheeler's superspace metric as a distance measure on the
space of mixmaster collapses (in the context of, for example, discussing
the chaotic properties) - instead of
using artificial and completely
arbitrary Euclidean distance measures \cite{Burd,Hobill,Rugh1990a}
on the solution space (which are not supported by the structure of general
relativity).

As concerns our discussion of the mixmaster gravitational collapse,
the (mini)superspace Ricci scalar $R$ may in principle be calculated
directly from the conformally rescaled Wheeler's (mini)superspace metric,
$$ \tilde{G}_{AB} = - e^{- 4 \Omega} (V(\beta_+, \beta_-) - 1)
\left(
\begin{array}{ccc}
-1&0&0\\
0&1&0\\
0&0&1
\end{array}
\right)
$$
However, ADM variables are not very suitable for such calculations
due to the complicated form of the potential term.
Since our aim is to extract the Ricci scalar $R(Maupertuis)$,
which is an invariant under canonical coordinate transformations,
it should not matter in which representation we work.
The ADM variables and the Bogoyavlenskii variables are related by
canonical coordinate transformations.\footnote{
More precisely,
one may transform the ADM variables to $\alpha, \beta, \gamma$
variables (cf. equation (\ref{ADMvariables})) and make a
corresponding canonical coordinate transformation of the
ADM momentum variables to $p_{\alpha}, p_{\beta}, p_{\gamma}$.
These are then the Bogoyavlenskii momentum variables,
cf. Bogoyavlenskii \cite{Bogoyavlenskii}, p. 40,
and one regains the Bogoyavlenskii
Hamiltonian from the ADM Hamiltonian up to a conformal factor
$a^2 b^2 c^2 = e^{2(\alpha + \beta + \gamma)}$ which is not important
because the Hamiltonians are equal to zero,
$H = 0$. }

Therefore, as in \cite{BieSzy91}, we rather
consider the Bogoyavlenskii Hamiltonian \cite{Bogoyavlenskii} in $(a,b,c)$
variables
\begin{eqnarray} \label{BogoyavlenskiiH}
 H & = & 2 (p_a p_b a^2 b^2 + p_b p_c b^2 c^2 + p_a p_c a^2 c^2) - p_a^2 a^4 -
p_b^2 b^4 - p_c^2 c^4 \nonumber  \\
 & + & \frac{1}{2} (a^2 b^2 + b^2 c^2 + a^2 c^2) - \frac{1}{4}
(a^4 + b^4 + c^4) \nonumber
\end{eqnarray}
(for the mixmaster collapse) in which case the potential is less
complicated at the expense that the metric read off from the kinetic energy is
no longer diagonal,
$$
a_{ij} =\frac{1}{2a^2b^2c^2} \left[
\begin{array}{ccc}
0 & c^2 & b^2 \\
c^2 & 0 & a^2 \\
b^2 & a^2 & 0
\end{array} \right].
$$
Explicit
evaluation of the Ricci scalar (\ref{RMaupertius}),
corresponding to the conformally rescaled
metric, induced by the kinetic term in the Bogoyavlenski Hamiltonian,
yields
\begin{equation} \label{RMaupertiusIX}
R_{IX} = - \frac{1}{8 (-V)^3} (a^8 + b^8 + c^8 - 2 a^4b^4 -
2 a^4c^4 - 2b^4c^4 + 16 a^2b^2c^2(a^2+b^2+c^2) )  \nonumber
\end{equation}
where
\begin{equation}
-V = + \frac{1}{2} g ({}^3 R) =
\frac{1}{4} (a^4 + b^4 + c^4 - 2 a^2b^2 - 2 a^2c^2 - 2 b^2c^2 )
\end{equation}

As concerns the aforementioned instability criterion applied to
the mixmaster three-geometries the idea is now to find regions
in the configuration space where $R < 0$ and
see whether
a typical trajectory is ``confined'' to
such regions or just traverse them quickly.
Clearly, $R$ cannot be negative all over
the entire minisuperspace. For example, it is easy to see
that $R>0$ in the case of isotropy,
$a = b = c.$ On the other hand
one observes
\cite{BieSzy91} that $R<0$ in the asymptotical
BKL regime, i.e. when $a >> b,c$ (or cyclic permutations thereof).

For sake of illustration we performed
a numerical experiment in which we have integrated the vacuum
mixmaster field equations
\cite{LandauLifshitzII}
\begin{eqnarray}
2 \alpha_{\tau \tau} & = &
\frac{d^2}{d \tau^2} (\ln a^2) = (b^2 - c^2)^2 - a^4 \nonumber \\
2 \beta_{\tau \tau} &= &
\frac{d^2}{d \tau^2} (\ln b^2) =
(c^2 - a^2)^2 - b^4 \\
2 \gamma_{\tau \tau} &=&
\frac{d^2}{d \tau^2} (\ln c^2) = (a^2 - b^2)^2 - c^4
\nonumber
\end{eqnarray}
supplemented by a first integral constraint
$$ I = \alpha_{\tau} \beta_{\tau} + \alpha_{\tau} \gamma_{\tau} +
\beta_{\tau} \gamma_{\tau} - \frac{1}{4}(a^4 + b^4 + c^4)
- 2 (a^2 b^2 - 2 a^2 c^2 - 2 b^2 c^2) = 0 \; . $$
Here $\tau = \int dt/abc$ denotes the standard time variable,
\cite{LandauLifshitzII} and subscript $\tau$
means differentiation with respect to the time variable $\tau$.
We used a numerical code from S.E. Rugh \cite{Rugh1990a} based on the
fourth-order Runge-Kutta integrator with check for the
first integral constraint (for details see \cite{Rugh1990a}).

The figure displays the temporal
(in $\tau$-time) behavior of the minisuperspace
Ricci scalar as felt by a trial trajectory.
We have selected a set of reference initial conditions as in
A. Zardecki \cite{Zardecki}
but have adjusted the value of $c'$ to make the first integral vanish
to {\em machine precision}.
(Such an adjustment is indeed  necessary. Cf. discussions in
S.E. Rugh \cite{Rugh1990a}  and D. Hobill \cite{Hobill}).
This yields the starting conditions
\begin{eqnarray} \label{startingconditions}
&a& = 1.85400.. \; \, \; b = 0.438500.. \; , \; c = 0.085400..  \nonumber \\
&a'&= -0.429200..  \; , \; b' = 0.135500.. \; , \;
c' = 2.964843279......
\end{eqnarray}
On the figure the minisuperspace Ricci scalar is displayed.
For comparison also the
evolution of (logarithmic) scale factors $\alpha, \beta, \gamma$ is
superimposed.
It turns out that $R$ is negative all the
time but varies over many orders of magnitude.
The latter is connected with the
singularities generated by a vanishing conformal factor
at points where the potential is zero. At these
points $R$ goes to (minus) infinity and a numerical cure against it was to
provide an arbitrary cut off.
This means that our trial trajectory of the mixmaster collapse stays in the
regions of minisuperspace where ``local instability'' occurs.
Of course such an illustration cannot be representative for any larger
set of trajectories. For example, the
minisuperspace Ricci scalar calculated
along a Taub axisymmetric solution was found to be positive.
However, even in cases when $R<0$ it is not possible
to claim, cf. our previous discussion about local and global properties,
that this fact accounts for global chaotic behavior - even though
some may be tempted to make such an inference.

\section{Concluding remarks.}

The possibility of mapping the dynamics of a
wide class of Hamiltonian systems
to geodesic flows on Riemannian manifolds
by application of Maupertuis principle,
and extracting coordinate invariant information about
local instability of the trajectories
(from the corresponding Riemann tensor)
has been applied in various contexts, ranging from N-body
systems (a gas) interacting via Van der Waals forces
(Krylov \cite{Krylov}), the instability (relaxation)
properties of a collisionless gas interacting via
gravitational forces (Gurzadyan and Savvidy \cite{GurzadyanSavvidy}) and
N-body systems interacting via the Debye-H\"{u}ckel potential
given by $V(r) \sim e^{-\kappa r}/r$ (a model for a
hot dilute plasma, Van Velsen \cite{vanVelsen}).

In this paper we have examined virtues and drawbacks of applying this
idea to general relativity \cite{BieSzy89}-\cite{SzyKraw}.
Coordinate invariant
information about local instability properties is
in this case
- as in the applications to the aforementioned
non-relativistic Hamiltonian systems - imagined to be extracted
either from negativity of the Ricci scalar (which is the
sum over principal sectional curvatures) or by
considering the (principal) sectional
curvatures themselves \cite{Bie94}.

We have shown that the Maupertuis principle as a way of geometrizing the
Hamiltonian dynamics when implemented to general relativity is actually
reproducing Wheeler's idea of a superspace in which
three-geometries evolve along geodesic lines. We have
examined this
superspace in order to get an insight into
the construction of gauge invariant (geometrical)
criteria for local instability
in general relativity. The original dream of having Einstein's
equations acting as a geodesic flow has been abandoned by its inventors a long
time ago because of artificial singularities created by the conformal
rescaling of the metric in points where the ADM potential equals zero. In so
far as for its original purpose i.e. to provide a suitable state-space for
quantization of gravity, people could (and actually did) retreat to more secure
ground of the weaker form of the superspace
(giving up the geodesic picture) the
{\em conformally rescaled}
superspace is crucial for our proposal. The idea is simple: If
one can cast the
dynamics into a geodesic flow then the sensitive dependence on initial
conditions (a key ingredient of instability and chaos)
may be captured at the level of Jacobi
equation for geodesic deviation.
The original intention was to point to a certain
reasonable approach rather than provide an ultimate solution (if
any exist) to the problem of invariant description of chaos. We have noticed
some obvious dangers when one makes global conclusions
from local considerations. This issue is far from being trivial however, since
the history of differential geometry teaches that starting from Gauss' {\em
theorema egregium} a lot of effort (sometimes spectacularly successful) has
been paid to the question of how to derive global characteristics from local
ones.

We have shown that expressions - used previously - such like
(\ref{RLyap}) relating directly the Ricci scalar
with Lyapunov exponents \cite{SzyLap,BieSzy91,SzSzczBie}
are not true. Therefore, unfortunately,
one cannot base a short time average (STA) Lyapunov exponent
on this expression either, cf. Burd and Tavakol \cite{BurdTavakol}.
The STA approach \cite{TavakolTvorkovski} as a tool for studying non-uniform
dynamics is based on
evaluation of the Lyapunov exponents (extracted in a standard way from the
Jacobian) over a finite time interval.
The mixmaster toy-model gravitational collapse examined here as
an example of a spacetime metric with complex (chaotic) behavior
is in fact rather well understood
---  the BKL-description \cite{BKL} is derived under assumptions which are very
good and indeed numerically confirmed in Rugh \cite{Rugh1990a}
and by Berger \cite{Berger1993}.
Nevertheless, if we want to
characterize - or rather to develop measures of - chaos in the
highly non-linear Einstein equations,
the mixmaster collapse is a suitable toy-model laboratory to
address such questions:
If we are not even able to invent useful
indicators of chaos in the moderately
simple example of the mixmaster space-time metric,
how are we ever going to be able
to deal with these issues for more general and more
complicated space-time metrics?

The observation of gauge variance
of standard indicators like that of a
Lyapunov exponent in the context of general relativity was
emphasized by Rugh \cite{Rugh1990a,Rugh1990b} and also
by Pullin \cite{Pullin}.
(See also recent discussion
in Rugh \cite{Rugh94}).
If we apply standard techniques of extracting the Lyapunov exponents
we at first sight have chaos in some gauges and in other gauges there
seems to be no chaos. Apparently, this situation
arises because we are looking at the
problem in the wrong way - we have not posed
a ``gauge invariant question''!

Whereas Rugh \cite{Rugh1990b} was
expressing hope of
constructing indicators
which are invariant under a
large class of coordinate transformations, i.e. indicators
which do not refer to a ``particular gauge'' (a problem which is easier
to point out than to solve) it appears that Pullin \cite{Pullin}
was resorting
to the Poincar\'{e} disc as a kind of ``selected set of coordinates''
--- a similar viewpoint is shared by Misner \cite{MisnerNATO}.
It appears to us
that no ``gauge'' is better than others
(is chaos in general relativity a concept which should only
be defined in certain selected frames of coordinates?)
- yet some gauges may make things particularly simple or
elegant thereby expressing the explicit or hidden symmetry (geometric
structure) of the problem. The geometric approach (via Maupertuis
principle) discussed in the present paper also points toward a certain
structure uncovered from the Hamiltonian formulation of general relativity
(which turned out to be equivalent to Wheeler's superspace) as a
``natural'' one. The problem is however not with ``preferred'' gauges but
rather
whether the criteria used to establish certain properties (such like
chaoticity) are gauge-invariant themselves.
Therefore we ought to seek measures of chaos (if they are
possible to construct) which are invariant under some large
class of gauge transformations.

Here we have examined an approach which points to a
more gauge invariant construction involving Wheeler's superpace.
However, this approach does (also) have several
obstacles, e.g. when implemented in the toy-model context of the mixmaster
gravitational collapse.
Hence it is safe to conclude that
how to characterize chaos most elegantly in
general relativity (even in the context of simple toy-models)
remains an open question.

\section*{Acknowledgements}

We would like to thank for discussions and fruitful exchanges
with Marek Szyd{\l}owski and with
Holger Bech Nielsen, Hans Henrik Rugh and Reza Tavakol.

M.B. wishes to thank the Niels Bohr Institute for warm
hospitality and support during his stays in Copenhagen. The support (of M.B.)
from the Foundation For Polish Science and from KBN Grant 20447 91 01 is
gratefully acknowledged. This work is also a contribution to the
KBN Grant 22108 91 02.

S.E.R. would like to thank support from
the Danish Natural Science Research Council, through grant
no. 11-8705-1.


\newpage

\section*{\sc Figure caption}

\begin{figure}
\caption[trialtrajectory]{
{\small
An illustration of the mixmaster minisuperspace Ricci scalar for a trial
trajectory.
Flat minima denote that the depth is out of range of the
figure. The Ricci scalar is negative all the time (but rapid changes
over many orders of magnitude makes it difficult to perceive this fact
on the figure). Indeed the artificial singularities introduced by the
conformally rescaled superspace metric (see the text)
make the Ricci scalar $R$ go to $- \infty$
while traversing the null-surface on which the potential is zero.
}
}
\end{figure}


\addcontentsline{toc}{section}{References}

\begin{thebibliography}{X}

\bibitem{MTW}
C.W. Misner, K.S. Thorne and J.A. Wheeler, {\em Gravitation}
(Freeman, New York, 1973).

\bibitem{ADM}
R. Arnowitt, S. Deser and C.W. Misner, {\em ``The Dynamics of
General Relativity''}
in {\em Gravitation, and Introduction to Current Research},
edited by L.Witten (Wiley, New York, 1962).


\bibitem{LandauLifshitzII}
L.D. Landau and E.M. Lifshitz, {\em The Classical Theory of
Fields}  (Pergamon, 1975).


\bibitem{EckmannRuelle}
J.P. Eckmann and D. Ruelle, {\em Rev.Mod.Phys.}~{\bf 57}, 617 (1985).

\bibitem{Lichtenberg}
A.J. Lichtenberg and M.A. Lieberman, {\em Regular and Stochastic Motion}
(Springer-Verlag, New York, 1983).

\bibitem{Benettin}
G. Benettin, L. Galgani and J.-M. Strelcyn,
{\em Phys.Rev.A}~{\bf 14}, 2338 (1976).

\bibitem{Ryan}
M.P. Ryan, {\em Hamiltonian Cosmology} (Springer-Verlag, 1972). \\
M.P. Ryan and L.C. Shepley, {\em Homogeneous Relativistic
Cosmologies} (Princeton University Press, 1975).

\bibitem{MacCallum}
M.A.H. MacCallum in {\em General Relativity -- An Einstein Centenary
Survey}, edited by S.W. Hawking and W. Israel
(Cambridge University Press, 1979).

\bibitem{Burd}
A. Burd, N. Buric and G.F.R. Ellis,
{\em Gen.Rel.Grav.}~{\bf 22}, 349 (1990).

\bibitem{Hobill}
D. Hobill, D. Bernstein, M. Welge and D. Simkins, in
{\em Proc.12th Int.Conf. on
General Relativity}, 337 (1990); {\em Class.Quant.Grav.}~{\bf 8}, 1155--1171
(1991).

\bibitem{Zardecki}
A. Zardecki, {\em Phys. Rev.D}~{\bf 28}, 1235 (1983).

\bibitem{Rugh1990a}
S.E. Rugh, {\em Chaotic Behavior and Oscillating Three-volumes in
a Space-Time Metric in General Relativity},
Cand. Scient. Thesis, The Niels Bohr Institute, Copenhagen
(January 1990). Second revised edition (April 1990). Distributed.
(Available upon request to the author).

\bibitem{RughJones}
S.E. Rugh and B.J.T. Jones, {\em Phys. Lett.A}~{\bf 147}, 353 (1990).

\bibitem{Rugh1990b}
S.E. Rugh  {\em ``Chaos in the Einstein Equations''}
Distributed at the Texas-ESO-CERN meeting at Brighton,
December 1990, Niels Bohr Institute preprint NBI-HE-{\bf 91-59} (1991).

\bibitem{Pullin}
J. Pullin, {\em ``Time and Chaos in General Relativity''}. Talk given at
the {\em VII SILARG Symposium}, Mexico City, December 1990.
Syracuse University preprint {\bf 90-0734} (1990).



\bibitem{RughJpn}
H.B. Nielsen and S.E. Rugh, {\em ``Chaos in the fundamental
forces ?''}, in Proc.Int.Symp.
``Quantum Physics and the Universe'', edited by
M. Namiki, {\em Vistas in Astronomy}~{\bf 37}, (1993).

\bibitem{Rugh94}
S.E Rugh {\em ``Chaos in the Einstein equations --- characterization and
importance?''}, 68 pp, Niels Bohr Institute preprint
NBI-HE-{\bf 94-07},
to appear in NATO ARW on ``Deterministic
Chaos in General Relativity'', edited by D. Hobill
(Plenum Press, New York, 1994).


\bibitem{BieSzy89}
M. Szyd{\l}owski and M. Biesiada, {\em Phys.Lett.B}~{\bf 220}, 32 (1989).

\bibitem{SzyLap}
M. Szyd{\l}owski and A. {\L}apeta, {\em Phys.Lett.A}~{\bf 148}, 239
(1990).

\bibitem{BieSzy91}
M. Szyd{\l}owski and M. Biesiada, {\em Phys.Rev.D}~{\bf 44},
2369 (1991).

\bibitem{SzSzczBie}
M. Szyd{\l}owski, J. Szcz{\c e}sny and M. Biesiada, {\em Chaos,Fractals and
Solitons}~{\bf 1}, 233 (1993).

\bibitem{SzyKraw}
M. Szyd{\l}owski and A. Krawiec, {\em Phys.Rev.D}~{\bf 47}, 5323 (1993).

\bibitem{Geneva}
M. Biesiada, {\em ``The Power of Maupertuis--Jacobi Principle''}
(1994) to appear in {\em Chaos, Fractals and Solitons}.

\bibitem{Bie94}
M. Biesiada,{\em ``Searching for invariant measures of chaos in
general relativity''},
submitted to {\em Class.Quant.Grav.} (1994).

\bibitem{Misner72}
C.W. Misner {\em ``Minisuperspace''} in {\em Magic Without Magic},
edited by J. Klauder (Freeman, San Francisco, 1972).


\bibitem{Bogoyavlenskii}
O. Bogoyavlenskii {\em Methods in the Qualitative Theory of Dynamical
Systems in Astrophysics and Gas Dynamics} (Springer, New York, 1985).

\bibitem{BurdTavakol}
A. Burd and R. Tavakol, {\em Phys.Rev.D}~{\bf 47}, 5336 (1993).

\bibitem{Krylov}
N.S. Krylov, {\em Works on the Foundations of Statistical Physics}
(Princeton University Press, New Jersey, 1979).

\bibitem{GurzadyanSavvidy}
V.G. Gurzadyan and G.K. Savvidy, {\em Astron. Astrophys.}~{\bf 160}, 203
(1986).

\bibitem{HedlundHopf}
G.A. Hedlund, {\em Bull.Amer.Math.Soc.}~{\bf 45}, 241 (1939)\\
E. Hopf, {\em Math.Ann.}~{\bf 117}, 590 (1940).


\bibitem{Chandrasekhar2}
S. Chandrasekhar, {\em Principles of stellar dynamics}
(University of Chicago Press, Chicago, 1943).

\bibitem{Lynden-Bell}
D. Lynden-Bell, {\em Mon.Not.R.Astr.Soc.}~{\bf 136}, 101 (1967).

\bibitem{Kandrup}
H. Kandrup, {\em Astrophys.J}~{\bf 364}, 420 (1990).

\bibitem{Arnold}
V.I. Arnold, {\em Mathematical Methods of Classical Mechanics}
(Springer Verlag, 1978).

\bibitem{vanVelsen}
J.F.C. van Velsen, {\em J.Phys.A:Math.Gen}~{\bf 13},
833 (1980).

\bibitem{Szebehely}
V. Szebehely, {\em Celestial Mechanics}~{\bf 34}, 49 (1984).


\bibitem{DahlqvistRussberg}
P. Dahlqvist and G. Russberg, {\em Phys. Rev. Lett.}
{\bf 65}, 2837 (1990).

\bibitem{Toda}
M. Toda, {\em Phys. Lett.A}~{\bf 48}, 335 (1974).

\bibitem{Brumer}
P. Brumer and J.W. Duff, {\em Journ. of Chem. Phys.}~{\bf 65}, 3566
(1976).

\bibitem{Tabor}
M. Tabor, {\em Adv. Chem. Phys.}~{\bf 46}, 73 (1981).


\bibitem{Teitelboim}
C. Teitelboim, {\em ``Hamiltonian Formulation of General Relativity''}
in {\em Quantum Cosmology and Baby Universes}, edited by
S. Coleman, J.B. Hartle, T. Piran and S. Weinberg (World
Scientific, Singapore, 1990).

\bibitem{DeWitt}
B. DeWitt, {\em Phys.Rev.}~{\bf 160}, 1113 (1967).

\bibitem{Uggla}
C. Uggla, K. Rosquist and R.T. Jantzen, {\em Phys.Rev.D}~{\bf 42}, 404
(1990).

\bibitem{BrownYork}
J.D. Brown and J.W. York, {\em Phys.Rev.D}~{\bf 47}, 1420 (1993).

\bibitem{TavakolTvorkovski}
R.K. Tavakol and A.S. Tvorkovski, {\em Phys.Lett.A},~{\bf 126}, 318
(1988).

\bibitem{BKL}
V.A. Belinskii, I.M. Khalatnikov and E.M. Lifshitz, {\em Sov.Phys. JETP}~{\bf
29}, 911 (1969);
{\em Adv.Phys.}~{\bf 19}, 525--573 (1970); {\em Adv.Phys.}~{\bf 31}, 639
(1982).

\bibitem{Barrow}
J.D. Barrow, {\em Phys.Rep.}~{\bf 85}, 1 (1982).

\bibitem{Berger1993}
B.K. Berger, {\em Phys.Rev.D}~{\bf 49}, 1120 (1994).


\bibitem{MisnerNATO}
C.W. Misner, {\em ``The Mixmaster Cosmological Metrics},
preprint, gr-qc/9405068, \\
to appear in NATO ARW on ``Deterministic
Chaos in General Relativity'', edited by D. Hobill
(Plenum Press, New York, 1994).



\end{thebibliography}
\end{document}